\documentclass[11pt, a4paper]{article}
\usepackage{jheparxiv}
\usepackage[latin1]{inputenc}
\usepackage{amsmath}
\usepackage{amsfonts}
\usepackage{amssymb}
\usepackage{latexsym}
\usepackage{mathrsfs}
\usepackage{graphicx}
\usepackage{color}
\usepackage{slashed}
\usepackage{twistor}
\usepackage[all]{xy}

\newcommand{\MT}{\mathbb{MT}}

\renewcommand{\d}{\mathrm{d}}

\subheader{\hfill \texttt{IMPERIAL-TP-TA-2017-04}}

\title{Minitwistors and 3d Yang-Mills-Higgs theory}

\author[a]{Tim Adamo,}
\author[b]{David Skinner}
\author[b]{\& Jack Williams}

\affiliation[a]{Theoretical Physics Group, Blackett Laboratory \\
        Imperial College London, SW7 2AZ, United Kingdom}

\affiliation[b]{Department of Applied Mathematics \& Theoretical Physics \\
        University of Cambridge, CB3 0WA, United Kingdom}

\emailAdd{t.adamo@imperial.ac.uk}
\emailAdd{[d.b.skinner, jw729]@damtp.cam.ac.uk}

\abstract{We construct a minitwistor action for Yang--Mills--Higgs theory in three dimensions. The Feynman diagrams of this action will construct perturbation theory around solutions of the Bogomolny equations in much the same way that MHV diagrams describe perturbation theory around the self--dual Yang Mills equations in four dimensions. We also provide a new formula for all tree amplitudes in YMH theory (and its maximally supersymmetric extension) in terms of degree $d$ maps to minitwistor space. We demonstrate its relationship to the RSVW formula in four dimensions and show that it generates the correct MHV amplitudes at $d=1$ and factorizes correctly in all channels for all degrees.}

\begin{document}
 
\maketitle

\section{Introduction}

This paper is concerned with perturbative aspects of Yang--Mills--Higgs theory 
\be\label{STLang1}
	S_0[A,\Phi]=-\frac{1}{2g^2}\int_{\R^3}\tr\left(F\wedge*F+\D\Phi\wedge*\D\Phi\right)\,,
\ee
in three dimensions.  In this action, $D=\rd+A$ is the covariant derivative and $F$ its curvature, the Higgs field $\Phi$ is a scalar in the adjoint of the gauge group, $*$ is the Hodge star on $\R^3$ with a flat Euclidean metric, and $g$ is a coupling constant. Note that $g^2$ has mass dimension $+1$ in $d=3$, so this theory is asymptotically free. This action is naturally interpreted as the dimensional reduction of pure Yang--Mills theory 
\be
	S[A^{(4)}] = -\frac{1}{2g_{(4)}^2}\int_{\R^3\times S^1} \tr\left(F^{(4)}\wedge*F^{(4)}\right)
\ee
on $\R^3\times S^1$ in the limit that the radius of the circle shrinks to zero size, with the Higgs field emerging as the component of $A^{(4)}$ along the $S^1$ directions, and where $g_{(4)}^2= {\rm Vol}(S^1)\,g^2$.

Many properties of YMH$_3$ are inherited from this relationship with YM$_4$. For the purposes of this paper, the key fact is that, just as it is possible to perturbatively expand YM$_4$ around the self--dual sector, so too YHM$_3$ admits a perturbative expansion around solutions of the Bogomolny equations~\cite{Ward:1985gz}
\be\label{Bog1}
	*F=\D\Phi\,.
\ee
These equations are the dimensional reduction of the self--duality condition $F^{(4)}= *^{(4)}F^{(4)}$ describing instantons in $d=4$. Solutions to~\eqref{Bog1} correspond to magnetic monopoles, and automatically solve the full field equations 
\be\label{STfeqs1}
	\D*F=[F,\Phi]\,, \qquad \D*\D\Phi=0
\ee
and Bianchi identity of~\eqref{STLang1}.

To expand YMH$_3$ around solutions of the Bogomolny equations, consider the new action
\be\label{STLnew1}
\begin{aligned}
	S[A,\Phi]&=-\frac{1}{2g^2}\int_{\R^3}\tr\left[(F-*\D\Phi)\wedge(*F-\D\Phi)\right]\\
	&=S_0[A,\Phi] +\frac{1}{g^2}\int_{\R^3} \tr\left(F\wedge D\Phi\right)\,.
\end{aligned}
\ee	
Using the Bianchi identity $DF=0$, the final term can be written $\int_{\R^3} {\rm d}\,\tr( F\Phi)$ and so is a total derivative that does not affect perturbation theory\footnote{In the presence of monopoles, this term is a topological invariant.} on $\R^3$. We now introduce a Lagrange multiplier $B$ which is a 1-form valued in the adjoint of the gauge group. With this field, the action \eqref{STLnew1} can be written as:
\be\label{CS1}
	S[A,B,\Phi]=\int_{\R^3}\tr\left[B\wedge(F-*\D\Phi)\right]+\frac{g^2}{2}\int_{\R^3}\tr\left(B\wedge*B\right)\,.
\ee
The equations of motion for this action are now
\be\label{CSeom}
\D\Phi-*F=g^{2}B\,, \qquad \text{and}\qquad \D B=-*[B,\Phi]\,,
\ee    
and it is easy to see that integrating out $B$ in the path integral results in the action \eqref{CS1}.

The significance of rewriting YMH$_3$ in the form~\eqref{CS1} is that the coupling $g$ now acts as a parameter for expanding around the monopole sector. Indeed, when $g=0$, $(A,\Phi)$ obey the Bogomolny equations while $B$ acts as a linear \emph{anti}-monopole gauge field propagating on the non-linear monopole background. For $g\neq0$, the field configuration is deformed away from the monopole equations by $*B$.  This construction is simply the dimensional reduction of the Chalmers-Siegel action~\cite{Chalmers:1996rq} for YM$_4$, which gives a perturbative expansion around the self-dual sector.

The monopole sector of YMH$_3$ is classically integrable (see {\it e.g.}~\cite{Nahm:1979yw,Ward:1981jb,Corrigan:1981fs,Hitchin:1982gh,Hitchin:1983ay,Murray:1985ji,Braden:2017ehl}), so has no non--trivial scattering (at least in Euclidean signature). This suggests that it should be an attractive background around which to study perturbation theory, just as the MHV expansion of YM$_4$ allows us to systematically construct $n$-particle gluon amplitudes, allowing arbitrary numbers of positive helicity $4d$ gluons at no cost~\cite{Cachazo:2004kj}.

\medskip

The aim of this paper is to study this perturbation theory, constructing all $n$-particle tree amplitudes in the theory~\eqref{CS1}. As in $d=4$, our approach will involve moving to twistor space where the integrability of the Bogomolny equations becomes manifest. For $\R^3$ the relevant twistor space is known as {\it minitwistor space}. This space was originally introduced by Hitchin in~\cite{Hitchin:1982gh} with the aim of studying the Bogomolny equations~\eqref{Bog1} (rather than full YMH$_3$ theory) and has been extensively studied since. We begin in section~\ref{sec:MT} by giving a brief introduction to the geometry of minitwistor space, largely following the perspective of~\cite{Jones:1984,Ward:1989vja}. We then review the Penrose transform and Hitchin--Ward correspondence, showing how solutions of the massless field equations and Bogomolny equations can be expressed in terms of cohomology classes and holomorphic vector bundles on minitwistor space. In section~\ref{sec:MTA} we construct an action in minitwistor space that describes YMH$_3$ theory perturbatively. We explicitly demonstrate the off--shell equivalence of this action with~\eqref{CS1}, and also show how it may be understood as the dimensional reduction of the twistor action for YM$_4$ found in~\cite{Mason:2005zm,Boels:2006ir}. We also consider the maximally supersymmetric version of the theory. Finally, we obtain a concise generating function for all tree--level amplitudes in (supersymmetric) YMH$_3$, written in terms of higher degree maps to minitwistor space. We show analytically that our expression for the amplitudes reproduces the correct 3--particle `$\overline{{\rm MHV}}$' and $n$-particle `MHV' amplitudes, and also that it factorizes correctly at all degrees. Finally, we explain that this generating function can be understood as the dimensional reduction of the RSVW expression~\cite{Witten:2003nn,Roiban:2004yf} for $\cN=4$ SYM in $d=4$.

Various parts of our story have appeared before. In particular, the minitwistor action corresponding to the `monopole' theory 
\be
	S[A,\Phi,B] = \int_{\R^3} \tr\left[B\wedge(F-*\D\Phi)\right]
\ee
({\it i.e.} the action~\eqref{CS1} at $g=0$) has appeared previously in~\cite{Popov:2005uv}, while expressions for the on--shell `MHV' {\it amplitudes} appeared in~\cite{Lipstein:2012kd}. However, the fact that the off--shell continuation of these amplitudes can be combined with the minitwistor action of~\cite{Popov:2005uv} to give an action for full YMH$_3$ appears not to have been appreciated. Amplitudes for YMH$_3$ have been studied in~\cite{Chiou:2005jn,Agarwal:2011tz,Cachazo:2013iaa}. In particular, in~\cite{Cachazo:2013iaa} Cachazo {\it et al.} gave a connected prescription formula for all tree amplitudes in YMH$_3$ that is equivalent to ours.  However, the formula of~\cite{Cachazo:2013iaa} was simply the $\cN=4$ SYM formula together with a set of $\delta$--functions enforcing that the external particles have no momentum in one of the four space--time directions. We show that the effects of these $\delta$--functions can be incorporated into a concise expression that is inherently three--dimensional. 

It is worth noting also that there is an alternative definition of twistors, given in~\cite{Huang:2010rn}, which are useful for describing amplitudes in 3d superconformal theoreies.


\section{Minitwistor Theory}
\label{sec:MT}

For our purposes, \emph{minitwistor space} will be the space of oriented geodesics in $\R^3$. In this section we review the geometry of minitwistor space and its supersymmetric generalisation, how free fields on $\R^3$ are encoded in minitwistor space, and the minitwistor description of the monopole sector.

\subsection{Geometry of minitwistor space}

Cartesian coordinates ${\bf x}$ on $\R^3$ may be encoded in a $2\times2$ symmetric matrix
\be\label{3dcoord}
	x^{\alpha\beta} = 
	\frac{\im}{\sqrt2} \begin{pmatrix}
					-x+\im y  & z\\
					z & x+\im y
				\end{pmatrix}
\ee						
where $\det{x^{\alpha\beta}} = \frac{1}{2}{\bf x}\cdot{\bf x}$ is (half) the Euclidean norm.

Minitwistor space $\MT$ is the total space of the holomorphic tangent bundle $T\CP^1$ to a Riemann sphere. As a line bundle, $T\CP^1\cong\cO(2)$, so we can describe $\MT$ using homogeneous coordinates $[u,\lambda_\alpha]$ where for $\alpha=0,1$, $\lambda_\alpha$ are homogeneous coordinates on the Riemann sphere $\CP^1$  and $u$ is a coordinate along the fibres at each point. These coordinates are considered up to overall $\C^*$ rescalings acting as
\be\label{mtscale}
	(u,\,\lambda_{\alpha})\sim (r^2 u,\,r\lambda_{\alpha})\,,
\ee
for all $r\in\C^*$, and the two $\lambda$s are never simultaneously zero.

The correspondence between $\MT$ and space--time is encoded in the {\it incidence relations}
\be\label{irels}
	u = x^{\alpha\beta}\lambda_\alpha\lambda_\beta\,,
\ee
where we temporarily allow the $x^{\alpha\beta}$ to be complex. For fixed $x^{\alpha\beta}$, equation \eqref{irels} describes a section $u: \CP^1\to T\CP^1$ so a point $x\in\C^3$ corresponds to a section of $\MT$ over $\CP^1$. We will call such sections {\it minitwistor lines}.  Note that {\it any} two minitwistor lines $X,Y\subset\MT$ (defined by $u=x^{\alpha\beta}\lambda_\alpha\lambda_\beta$ and $u=y^{\alpha\beta}\lambda_\alpha\lambda_\beta$) will intersect each other in two points, since $(x-y)^{\alpha\beta}\lambda_\alpha\lambda_\beta=0$ can be regarded as a quadratic equation in the local coordinate $z = \lambda_1/\lambda_0$. This reflects the fact that the normal bundle to each minitwistor line is $\cO(2)$. 

Suppose we label the two intersection points in $\MT$ by $(u,\lambda_{\alpha})$ and $(u',\lambda_{\alpha}^{\prime})$. The incidence relations imply that the vector connecting the two points in $\R^3$ must take the form $(x-y)^{\alpha\beta}=\lambda^{(\alpha}\,\lambda^{\prime\,\beta)}$. Hitchin~\cite{Hitchin:1982gh} defines a (holomorphic) conformal structure on $\C^3$ by declaring $x,y\in\C^3$ to be null separated iff the discriminant of this quadratic vanishes, so that the two intersection points $X\cap Y$ coincide. 

In the other direction, for a fixed point $[u,\lambda_\alpha]\in\MT$, given one point $x_0\in\C^3$ obeying~\eqref{irels} we can construct the two--parameter family
\be\label{alphaplane}
	x^{\alpha\beta}(\nu) = x_0^{\alpha\beta} + \nu^{(\alpha}\lambda^{\beta)}
\ee
which also obeys~\eqref{irels} for any choice of $\nu^\alpha$. Thus, a point in $\MT$ corresponds to a totally null complex 2-plane $\C^2\subset\C^3$. 

\medskip

To consider real Euclidean $\R^3$, rather than $\C^3$, we impose a reality condition on our sections. As in twistor space for four dimensions, consider the antiholomorphic involution 
$\MT\to\MT$ defined by
\be\label{reality0}
	u\mapsto\hat{u} = \bar u\,,\qquad\lambda_{\alpha}\mapsto\hat{\lambda}_\alpha = (-\bar\lambda_1,\bar\lambda_0)\,.
\ee
This antiholomorphic involution acts on the $\CP^1$ as the antipodal map, and so has no fixed points. However, there are fixed minitwistor lines. We have
\be\label{reality1}
	\hat{x}^{\alpha\beta}=-\frac{\im}{\sqrt{2}}
	\begin{pmatrix}
	 \bar x - \im \bar y & -\bar z \\
	 -\bar z & - \bar x - \im \bar y
	 \end{pmatrix}\,,
\ee
so demanding that $\hat{x}^{\alpha\beta} = x^{\alpha\beta}$ imposes $(x,y,z)\in\R^3$ as desired. If $x_0\in\R^3$ is real Euclidean, then the $\C^2\subset\C^3$ given by~\eqref{alphaplane} intersects the real Euclidean slice only where $\nu_\alpha = \im r\hat{\lambda}_\alpha$ for some $r\in\R$. Thus, in Euclidean signature, a point in $\MT$ corresponds to a straight line in $\R^3$.  

\begin{figure}[t]
\centering
\includegraphics[width=110mm]{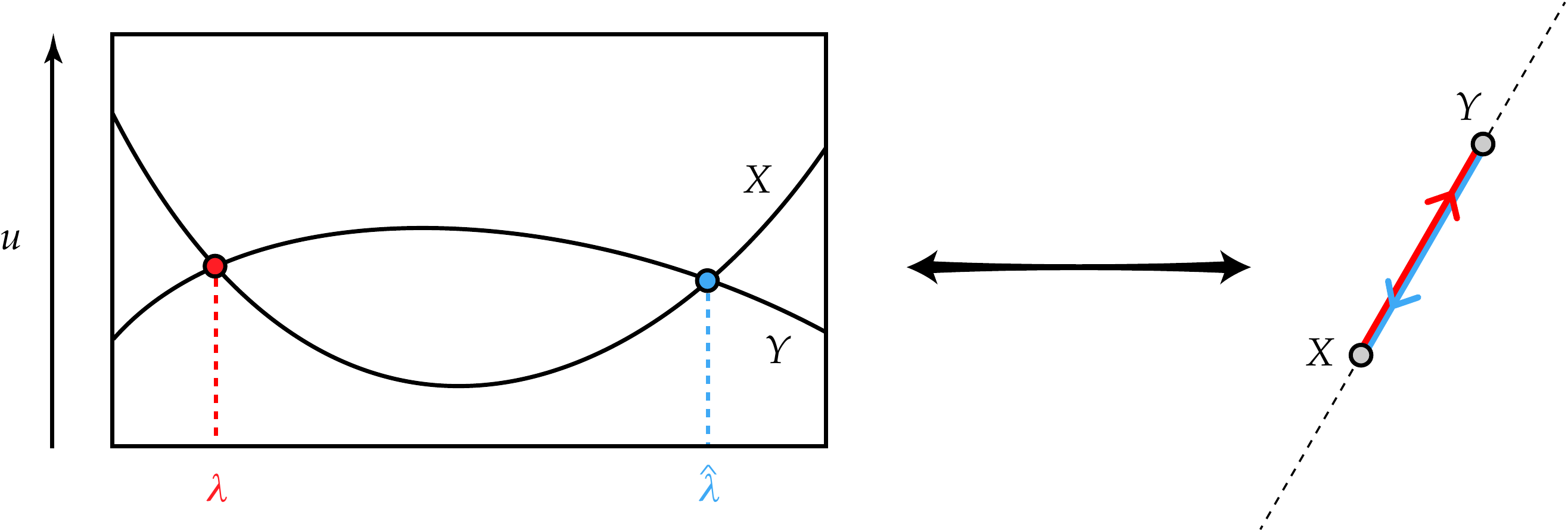}
\caption{\emph{The correspondence between $\MT$ (right) and $\R^3$ (left).}}
\label{mtcorr}
\end{figure}

The minitwistor lines of two Euclidean real points still intersect twice in $\MT$; the two intersection points correspond to the two (opposite) orientations of the unique geodesic in $\R^3$ which connects the two points. See Figure~\ref{mtcorr}. If one of the intersection points of the two minitwistor lines is $(u,\lambda_{\alpha})$, the norm of the connecting vector $(x-y)^{\alpha\beta}$ is proportional to $\la\lambda\hat{\lambda}\ra^2$. This norm is invariant under the Euclidean reality conditions, as required.

So a point in $\R^3$ corresponds to a minitwistor line in $\MT$, while a point in $\MT$ corresponds to a geodesic (i.e., a straight line) in $\R^3$.

Altogether, the correspondence is summarised by the diagram 
\begin{equation*}
\xymatrix{
 & \PS \ar[ld]_{\pi_1} \ar[rd]^{\pi_2} & \\
 \MT & & \R^3 }
\end{equation*}
where $\PS\cong\R^3\times\CP^1$ is the projective spinor bundle with coordinates $(x^{\alpha\beta},[\lambda_\gamma])$. The projection $\pi_{1}:\PS\rightarrow\MT$ is given by the incidence relations discussed above, while the second fibration $\pi_2$ is the trivial projection $(x,\lambda)\mapsto x$.

\medskip

For our purposes it will be useful to note that, just as $\R^3$ can be obtained by taking the quotient of $\R^4$ along a constant vector field, so too can $\MT$ be obtained by taking the quotient of the twistor space of $\R^4$ by the action of this vector field on $\PT$~\cite{Hitchin:1982gh,Jones:1984,Ward:1989vja}. Explicitly, let $\R^4$ have coordinates $x^{\alpha\dot\alpha}$ and let $T = T^{\alpha\dot\alpha} \del/\del x^{\alpha\dot\alpha}$ denote a constant vector field. In the coordinates
\be
	x^{\alpha\dot\alpha} = \frac{\im}{\sqrt{2}}
	\begin{pmatrix}
			-x+\im y & z + t\\
			z-t & x+\im y
	\end{pmatrix}
	\qquad\text{we have}\qquad
	T^{\alpha\dot\alpha} = 
	\begin{pmatrix}
			0 &\ 1\ \\
			-1 &\ 0\
	\end{pmatrix}\,.
\ee
It follows that the three- and four-dimensional coordinates are related as $x^{\alpha\beta} = x^{(\alpha}_{\ \dot\alpha} T^{\beta)\dot\alpha}$. The twistor space of $\R^4$ is the total space of the rank-2 bundle $\cO(1)\oplus\cO(1)\to\CP^1$, and is called $\PT$. In terms of homogeneous coordinates $[Z^{A}]=[\mu^{\dot{\alpha}},\lambda_{\alpha}]$ on $\PT$, the incidence relations are $\mu^{\dot\alpha} = x^{\alpha\dot\alpha}\lambda_{\alpha}$. The vector $T$ defining the symmetry reduction $\R^4\rightarrow\R^3$ acts on $\PT$ as
\be\label{PTvec}
	T=T^{\alpha\dot\alpha}\,\lambda_{\alpha}\,\frac{\partial}{\partial\mu^{\dot\alpha}}\,.
\ee
Minitwistor space is then obtained by factoring out $\PT$ by the integral curves of this vector field; in particular, the minitwistor coordinate $u$ can be written in terms of coordinates on $\PT$ as $u = \mu_{\dot\alpha}\, T^{\alpha \dot\alpha}\,\lambda_\alpha$, which is annihilated by~\eqref{PTvec} and naturally has scaling weight $+2$. The real structure on $\MT$ responsible for Euclidean reality conditions on space--time are inherited from a similar real structure on $\PT$ (c.f., \cite{Woodhouse:1985id}).

\medskip

We will also be interested in a supersymmetric extension of $\MT$, denoted $\MT_s$. For even $\cN$ this is straightforward to achieve: as in~\cite{Popov:2005uv} we promote $\MT$  to the total space of the bundle
\be\label{smt1}
	\cO(2)\oplus \left(\C^{0|\frac{\cN}{2}}\otimes\cO(1)\right)\rightarrow \CP^1\,,
\ee
which can be described by homogeneous coordinates $[u,\chi^a,\lambda_{\alpha}]$ where the $\chi^{a}$ are Grassmann variables and $a=1,\ldots,\cN/2$. These coordinates are considered up to the rescaling
\be\label{smt2}
	(u,\,\lambda_{\alpha},\,\chi^{a}) \sim (r^2 u,\,r\chi^{a},\,r\lambda_{\alpha})\,,
\ee
for any $r\in\C^*$.  Note that only half the full supersymmetry will be manifest in this description. The incidence relations are similarly generalised to
\be\label{smtinc}
	u=x^{\alpha\beta}\,\lambda_\alpha \lambda_{\beta}\,, \qquad \chi^{a}=\theta^{a\alpha}\,\lambda_{\alpha}\,,
\ee
describing a section of~\eqref{smt1}, where $\theta^{a\alpha}$ are coordinates on chiral superspace, written in an $\SU(4)$-invariant formalism. Euclidean reality conditions on the Grassmann variables on $\MT_s$ are inherited from those for supersymmetric extensions of $\PT$~\cite{Boels:2006ir}. For example, for $\cN=8$ the conjugation \eqref{reality0} is extended to:
\be\label{realitys}
\chi^{a}\mapsto\hat{\chi}^{a}=\left(-\bar{\chi}^{2},\,\bar{\chi}^{1},\,-\bar{\chi}^{4},\,\bar{\chi}^3\right)\,,
\ee
and real minitwistor lines in $\MT_s$ are preserved under this conjugation.

We note that when $\cN=8$ (so $\cN/2=4$) the Berezinian\footnote{The Berezinian is the supermanifold analogue of the canonical line bundle.} line bundle has a global holomorphic section given by
\be\label{smtmeas}
	\Omega_{\MT_s} \equiv \frac{\d u \wedge\d^{2|4}\lambda}{\mathrm{vol}\,\C^*}=\d u\wedge \la\lambda\,\d\lambda\ra\  \d^{0|4}\chi\,.
\ee
This has weight zero under the projective rescalings \eqref{smt2}. Thus $\cN=8$ minitwistor space is a Calabi--Yau supermanifold. We also note that $\Omega_{\MT_s} = T\lrcorner \D^{3|4}Z$ where $\D^{3|4}Z$ is a global holomorphic section of the trivial Berezianian of $\cN=4$ twistor space $\PT_s$.



\subsection{Penrose transform \& Hitchin-Ward correspondence}

The Penrose transform represents zero-rest-mass free fields in terms of cohomological data on twistor space~\cite{Penrose:1969ae,Eastwood:1981jy}. A  version of this transform exists for minitwistor space, where it has an interesting interaction with the linearised Bogomolny monopole equations~\cite{Jones:1984,Ward:1989vja,Tsai:1996}. We now review the relevant aspects of the mintwistor Penrose transform.

\medskip

Let $f\in \Omega^{0,1}(\MT,\cO(-n-2))$ be a (0,1)-form on minitwistor space which is homogeneous of weight $-n-2$. If $n\geq0$ we can define a spin $n/2$ field on $\R^3$ from this $f$ using the integral transform
\be\label{PT1}
\varphi_{\alpha_{1}\cdots\alpha_{n}}(x)=\int_{X}\la\lambda\,\d\lambda\ra\wedge \lambda_{\alpha_1}\cdots\lambda_{\alpha_n}\,f|_{X}\,,
\ee
where $f|_X$ indicates restricting the arguments of $f$ to the minitwistor line $X\subset\MT$ via the incidence relations \eqref{irels}. Provided $f$ obeys $\dbar f=0$, so that $f$ is holomorphic on $\MT$, it follows that $f|_{X}=f(x^{\gamma\delta}\lambda_{\gamma}\lambda_{\delta}, \lambda_{\alpha})$ and the incidence relations ensure that $\varphi_{\alpha_1 \cdots\alpha_n}$ obeys the spin $n/2$ free field equations on $\R^3$:
\be\label{PT2}
\partial^{\alpha_{1}\beta}\varphi_{\alpha_{1}\cdots\alpha_{n}}(x)=\int_{X}\la\lambda\,\d\lambda\ra\wedge \lambda_{\alpha_1}\cdots\lambda_{\alpha_n}\,\lambda^{\alpha_{1}}\lambda^{\beta}\left.\frac{\partial f}{\partial u}\right|_{X} = 0\,.
\ee
On the other hand, if $f = \dbar g$ for some $g\in\Omega^0(\MT, \cO(-n-2))$ then the spinor field~\eqref{PT1} is identically zero. Thus, non--trivial solutions to the spin $n/2$ free--field equations correspond to elements of the cohomology class $H^{0,1}(\MT,\cO(-n-2))$ and it can be shown that this relation is in fact an isomorphism~\cite{Jones:1984,Tsai:1996}:
\be\label{PT4}
	H^{0,1}(\MT,\,\cO(-n-2))\cong\left\{\mbox{spin } \frac{n}{2} \mbox{ zero-rest-mass fields on } \R^3\right\}\,,
\ee
provided the set of zero-rest-mass fields is subject to appropriate analyticity conditions. When $n=0$ we have  
\be\label{PT3}
	f\in H^{0,1}(\MT,\cO(-2))\,, \qquad \varphi(x)=\int_{X}\la\lambda\,\d\lambda\ra\wedge f|_{X}\,,
\ee
corresponding to a (complex) solution of the scalar wave equation $\Box\varphi=0$ on $\R^3$.

When $n<0$, the integral transform must be altered. Given a representative $\psi\in H^{0,1}(\MT,\cO(-1))$, one can still define a spin $1/2$ zero-rest-mass field by taking
\be\label{PT*1}
\Psi_{\alpha}(x)=\int_{X}\la\lambda\,\d\lambda\ra\wedge \lambda_{\alpha}\,\left.\frac{\partial \psi}{\partial u}\right|_{X}\,,
\ee
with $\partial^{\alpha\beta}\Psi_{\alpha}=0$ following by the same argument as before. Note that in three dimensions, this spinor has the same Weyl index as that constructed from $f\in H^{0,1}(\MT,\cO(-3))$.  However, something new occurs for minitwistor fields of weight zero. Given a representative $a\in H^{0,1}(\MT,\cO)$, we can define both a scalar and a spin-one field on $\R^3$ by the integral transforms
\be\label{PT*2}
	\Phi(x)=\int_{X}\la\lambda\,\d\lambda\ra\wedge\left.\frac{\partial a}{\partial u}\right|_{X} \qquad \text{and}\qquad (*f)_{\alpha\beta}(x)=\int_{X}\la\lambda\,\d\lambda\ra\wedge\lambda_{\alpha}\lambda_{\beta}\,\left.\frac{\partial^{2}a}{\partial u^2}\right|_{X}\,,
\ee
where we have chosen to write the spin-one field as the dual of a linearised field strength. It is easy to see these fields each obey the relevant field equation $\Box\Phi=0$ or $\partial^{\alpha\gamma} (*f)_{\alpha\beta}=0$. However, since they are built from the same twistor field $a$, one expects that $\Phi$ and $*f$ are not independent. In fact, we have
\be\label{linBM}
	\partial_{\alpha\beta}\Phi=\int_X\la\lambda\,\d\lambda\ra \wedge \lambda_\alpha\lambda_\beta \left.\frac{\del^2 a}{\del u^2}\right|_X  
	= (*f)_{\alpha\beta}\,,
\ee
showing that a minitwistor representative $a\in H^{0,1}(\MT\,\cO)$ encodes both a massless scalar and a Maxwell field which are related by the Bogomolny equations. 

\medskip

The Penrose transform also allows us to encode the full field content of $\cN=8$ SYMH theory into a single field on supersymmetric minitwistor space $\MT_s$. If $\cA\in H^{0,1}(\MT_s,\cO)$, then $\cA$ can be expanded in the fermionic directions  as
\be\label{sPT1}
	\cA=a+\chi^{a}\psi_{a}+\frac{1}{2}\chi^{a}\chi^{b}\,\varphi_{ab}+\frac{1}{3!}\epsilon_{abcd}\,\chi^{a}\chi^{b}\chi^{c}\,\tilde{\psi}^d+\frac{\chi^4}{4!}b\,,
\ee
with the individual $(0,1)$-forms $\{a,\psi_{a}, \varphi_{ab},\tilde{\psi}^{a},b\}$ on bosonic minitwistor space having weights $0,-1,\ldots,-4$, respectively. Under the Penrose transform, the component fields $\varphi_{ab}$ correspond to 6 space-time scalars, the fields $\psi_a$ and $\tilde\psi^a$ together yield 8 Weyl fermions, whilst $a$ and $b$ together describe both a linearised gluon and a further scalar. As above, the combination that solves the linearised Bogomolny equations is contained in $a$, whilst $b$ describes a solution to the anti--Bogomolny equations. 

This is the dimensional reduction of the statement that on $\PT$, the weight 0 field corresponds to a positive helicity (self--dual) field on $\R^4$, whilst the weight $-4$ field corresponds to a  negative helicity (anti--self--dual) field. Altogether, $\cA$ in~\eqref{sPT1} describes the linearised field content of maximal ($\cN=8$) SYMH in three dimensions, in a framework where only an $\SU(4)$ subgroup of the full SO$(8)$ $R$-symmetry is manifest. This subgroup is fixed by choosing which of the seven scalars should be paired with the gauge field in the Bogomolny equations.

\medskip

At the non-linear level, the \emph{Hitchin--Ward correspondence}~\cite{Ward:1981jb,Hitchin:1982gh} describes solutions of the full Bogomolny equations on $\R^3$ in terms of holomorphic vector bundles on minitwistor space. This correspondence is inherited from the Ward construction of  instantons in YM$_4$ via holomorphic vector bundles over the twistor space $\PT$~\cite{Ward:1977ta}. The Hitchin--Ward construction is equivalent to other well-known constructions of monopoles~\cite{Hitchin:1983ay}, such as the Nahm equations~\cite{Nahm:1979yw}.  The precise statement of the Hitchin-Ward correspondence is as follows~\cite{Hitchin:1982gh}. There is a one-to-one correspondence between $\SU(N)$ Bogomolny monopole configurations on $\R^3$ and rank $N$ holomorphic vector bundles $E\rightarrow\MT$ which obey {\it i)} $E|_{X}$ is topologically trivial for every minitwistor line $X\subset\MT$, {\it ii)} $\det E$ is trivial and {\it iii)} $E$ admits a positive real form. The latter two conditions are related to the choice of special unitary gauge group\footnote{Triviality of $\det E$ ensures the existence of a nowhere vanishing holomorphic section of $\det E$ which can be used to normalize the transition matrices of $E$ to have unit determinant. A positive real form on $E$ defines the Killing form on $\SU(N)$.} and the correspondence can be generalised to any choice of gauge group. For our purposes, the most important feature is that holomorphic bundles on $\MT$ correspond to general solutions of the Bogomolny equations on $\R^3$.


\section{The Minitwistor Action}
\label{sec:MTA}

In this section, we reformulate YMH$_3$ in terms of minitwistor data. This follows by translating the space-time action \eqref{CS1} into a `minitwistor action.'  As we shall see, this action is actually defined on the projective spinor bundle $\PS$, but the equations of motion are naturally phrased in terms of minitwistor space. After formulating the minitwistor action for $\cN=0$ and $\cN=8$ and showing that it has the appropriate equations of motion, we also prove that it reduces to the action on $\R^3$ with a particular choice of gauge.

This construction closely parallels that of the twistor action for Yang-Mills theory in four-dimensions~\cite{Mason:2005zm,Boels:2006ir,Adamo:2013cra}, and shares many of its features.


\subsection{Action and equations of motion}

In the introduction to this paper we saw that YMH$_3$ admits a perturbative expansion around the Bogomolny monopole sector. This was apparent by writing the action as
\be\label{stact1}
S[A,B,\Phi]=S_{\rm m}[A,B,\Phi]+\frac{g^2}{2}\,I[B]\,,
\ee
where 
\be\label{stactm}
	S_{\rm m}[A,B,\Phi] = \int_{\R^3} \tr\left[B \wedge(F - *\D\Phi)\right]
\ee
describes the monopole sector and
\be\label{stacti}
	I[B] = \frac{g^2}{2}\int_{\R^3} \tr\left(B\wedge*B\right)
\ee
deforms the monopole equations to the full YMH equations. Weak coupling $g=0$ corresponds to the monopole sector itself.

By the Hitchin--Ward correspondence, the Bogomolny equations $F = *\D\Phi$ are equivalent to holomorphic bundles over $\MT$. Thus, a first attempt at constructing a minitwistor version of $S_{\mathrm{m}}$ might be to simply impose these holomorphicity conditions directly on $\MT$ by means of a Lagrange multiplier~\cite{Popov:2005uv}. Let $E\rightarrow\MT$ be a rank $N$ complex (but not necessarily holomorphic) vector bundle, and assume\footnote{As in the Hitchin--Ward correspondence, for an SU$(N)$ theory we must also assume that $\det E$ is trivial and that $E$ has a positive real form.} that $E$ is trivial on restriction to the holomorphic lines $X\subset\MT$ corresponding to points $x\in\R^3$. We can endow $E$ with a partial connection $\bar{D}$, locally of the form $\bar{D}=\dbar+a$ for $a\in\Omega^{0,1}(\MT,\cO\otimes\End(E))$. In other words, $a$ is the $(0,1)$-gauge potential for the partial connection $\bar{D}$. If $F^{0,2}=\bar{D}^2$ is the curvature of $\bar{D}$, then a first guess at a minitwistor action for the monopole sector could take the form
\be\label{MTmono1}
	S_{\mathrm{m}}[a,\beta]\buildrel{?}\over{=}\int_{\MT}\Omega\wedge\tr\!\left(\beta\,F^{0,2}\right)\,,
\ee
where $\Omega\equiv \d u\wedge\la\lambda\,\d\lambda\ra$ is the top holomorphic form of weight $+4$ on $\MT$ and $\beta$ is an $\End(E)$-valued function on $\MT$ of weight $-4$. Varying $\beta$, one obtains the equation of motion $F^{0,2}=0$; by the Hitchin--Ward correspondence, this corresponds to a gauge field $A$ and scalar $\Phi$ on $\R^3$ obeying the monopole equations. 

Varying $a$, the other equation of motion we obtain from this action is $\bar{D}\beta=0$, which imposes that $\beta$ is globally holomorphic with respect to the complex structure $\bar{D}$. However, the space $H^{0}(\MT,\cO(-4))$ of solutions to this equation is in fact empty, as one can see {\it e.g.} by restricting $\beta$ to any minitwistor line, where it would have to be a globally holomorphic function of weight $-4$ on the Riemann sphere. Thus the second equation implies that $\beta=0$ and so encodes no physical degrees of freedom on $\R^3$. By contrast, the space-time equations 
\be
	\D *B =0 \qquad\text{and} \qquad \D B = -*[B,\Phi]
\ee
following from varying~\eqref{stactm} with respect to $\Phi$ {\it do} have non--trivial solutions, corresponding to linearised anti-Bogomolny fluctuations around the monopole. Thus~\eqref{MTmono1} cannot be the minitwistor action corresponding to~\eqref{stactm}.

\medskip

The remedy for this is perhaps surprising. Rather than constructing an action on $\MT$ itself, we instead consider an action on the projective spinor bundle $\PS$. We now let $E\to\PS$ be a vector bundle over $\PS$ and consider the action\footnote{We abuse notation by writing the pullback $\pi_1^*\Omega$ to $\PS$ of the top form on $\MT$ also as $\Omega$.}
\be\label{MTmono}
	S_{\mathrm{m}}[a,b]=\int_{\PS}\Omega\wedge\tr\!\left(b\wedge\cF\right)\,,
\ee
where now $a$ and $b$ are ${\rm End}(E)$-valued 1-forms on $\PS$ of weight zero and $-4$, respectively, and $\cF = \d a+a\wedge a$ is the curvature of $\d+a$. The action is invariant under the gauge transformations
\be\label{gi1}
	\d+a\rightarrow g\left(\d +a\right)g^{-1}\,, \qquad b\rightarrow g\, b\,g^{-1}\,,
\ee
for $g\in\Omega^0(\PS,{\rm End}(E))$, and also under the shift transformation
\be\label{gi2}
	a\rightarrow a\,,\qquad b\rightarrow b+(\d+a)\beta\,,
\ee
for $\beta$ an $\End(E)$-valued function of weight $-4$ on $\PS$. The latter transformation follows from the Bianchi identity for $\cF$, and is standard in $BF$ theories.

\emph{A priori}, this action may seem a long way from what we are hoping for as both the action and variational data are defined on $\PS$ rather than $\MT$. The equations of motion following from~\eqref{MTmono} state that
\be
	\Omega\wedge\cF = 0 \qquad\text{and} \qquad \Omega\wedge\left( \d b+ [a,b]\right)=0\,.
\ee
We can analyse the content of these equations as follows. First note that $\PS\cong\MT\times\R$, so all differential forms on $\PS$ can be expanded in a basis of forms on $\MT$ and the real fibre. It is clear that the action~\eqref{MTmono} only depends on the components of $a$, $b$ that span the antiholomorphic directions of $\PS$ together with the fibre direction, since the other components wedge to zero against $\Omega$. For the variational problem encoded in this action, one can therefore expand
\be\label{vardat}
	a=\bar {a} + a_\perp\,, \qquad b=\bar{b} + b_{\perp}\,,
\ee
where $a_\perp$, $b_\perp$ represent 1-forms on $\PS$ pointing along the fibre of the projection $\PS\to\MT$, while $\bar{a}$ and $\bar{b}$ represent forms on $\PS$ that point in the antiholomorphic directions of $\MT$. We can similarly decompose
\be
	\d = \del + \dbar + \d_\perp
\ee
and note that the presence of $\Omega$ again means that the exterior derivative $\del$ in the holomorphic directions of $\MT$ drops out.  

We now use \eqref{gi1} and \eqref{gi2} to set $a_{\perp}=0=b_{\perp}$. As always with axial gauges, this condition does not completely fix the gauge and we are still free to make gauge transformations that are independent of the fibre coordinates -- {\it i.e.} we can still perform gauge transformations on $\PS$ that are the pullback of smooth gauge transforms on $\MT$. In this gauge the equations of motion for $\bar{a}$ read
\be\label{aeom}
	\cF^{0,2} = 0\qquad\text{and}\qquad\d_{\perp} \bar{a} = 0
\ee
where $\cF^{0,2} = \dbar \bar{a} +\bar{a}\wedge\bar{a}$, while the equations of motion for $\bar b$ become
\be\label{beom}
	\dbar \bar{b} + [\bar{a},\bar{b}] = 0\qquad\text{and}\qquad \d_\perp \bar{b}=0\,.
\ee
The equations $\d_\perp{\bar a}=0=\d_\perp\bar{b}$ tell us that the remaining components $\bar{a}$ and $\bar{b}$ are independent of the real fibre coordinate, so on--shell these $\bar{a}$ and $\bar{b}$ are in fact $(0,1)$-forms on $\MT$, pulled back to $\PS$. The remaining equations $\cF^{0,2}=0$ say that on--shell, the bundle $E\to\PS$ is just the pullback of a holomorphic bundle on $\MT$, while the equation $\bar{D}\bar{b}=0$ together with the residual gauge invariance says that $\bar{b}$ represents an element of $H^{0,1}(\MT,{\rm End}(E)\otimes\cO(-4))$, pulled back to $\PS$. These are exactly the desired equations on $\MT$, corresponding by the Hitchin--Ward correspondence and the covariant extension of the linear Penrose transform~\eqref{PT4} to the equations
\be
	F -*\D\Phi =0 \qquad\text{and}\qquad \D B = -*[B,\Phi]
\ee
on $\R^3$. Thus, at least on--shell, the action~\eqref{MTmono} corresponds to the action~\eqref{stactm} on $\R^3$. (We note that both actions vanish trivially when evaluated on a solution of their equations of motion.)

\medskip

It remains to define the anti-monopole interaction term \eqref{stacti} in terms of our fields. In the case of an abelian gauge group, it is clear how to proceed thanks to the Penrose transform. Indeed, at least on--shell we have
\be\label{MTint1}
	\int_{\R^3} B\wedge *B
	=\int\limits_{\R^{3}\times\CP^1\times\CP^1} \d^{3}x\,\la\lambda_1 \d\lambda_{1}\ra\,\la\lambda_{2} \d\lambda_{2}\ra\, \la\lambda_1 \lambda_2\ra^2\,
	\bar{b}(x,\lambda_1)\, \bar{b}(x,\lambda_2)\,
\ee
in the abelian case, where $\bar{b}$ is as before. 

In the non-abelian case, where $b$ takes values in $\End(E)$, we must modify this term. Since our bundle $E\to\PS$ was assumed to be trivial on restriction to each minitwistor line, on any given minitwistor line there is a smooth gauge transform $h \in \Omega^0(X,{\rm End}(E))$ such that
\be\label{htriv1}
	h(x,\lambda)\,\bar{D}|_{X}\,h^{-1}(x,\lambda)=\dbar|_{X}.
\ee
Clearly, such an $h$ exists thoughout $\PS$ when $a=0$, and so will continue to exist for $a$ sufficiently small. Thus, in perturbation theory, a holomorphic trivialization $h$ will always exist. Furthermore, since $X\subset\MT$ is linearly embedded (via the incidence relations), this holomorphic trivialization is unique up to multiplication $h \mapsto h h_0$ where $h_0$ is independent of $\lambda$.  This allows for a \emph{non-abelian} definition of the integral formulae for the Penrose transform (c.f., \cite{Mason:2005zm}), with
\be\label{htriv2}
B_{\alpha\beta}(x)=\int_{X}\la\lambda\,\d\lambda\ra\  \lambda_{\alpha}\lambda_{\beta}\,h\,\bar{b}|_{X}\,h^{-1}\,,
\ee
leading to a 1-form on $\R^3$ also valued in the adjoint of the gauge group.

The holomorphic trivialization $h$ can be used to define holomorphic frames~\cite{Mason:2010yk,Bullimore:2011ni}
\be\label{htriv3}
U_{X}(\lambda,\lambda')\equiv h(x,\lambda)\,h^{-1}(x,\lambda')\,,
\ee
which map the fibre of $E|_{X}$ at $\lambda'$ to the fibre at $\lambda$. By definition, 
\be
	\bar{D}|_{X} U_{X}=0\qquad\text{and}\qquad U_{X}(\lambda,\lambda)=\mathbf{1}\in\End(E)\,.
\ee
The non-abelian generalisation of \eqref{MTint1} is therefore
\begin{multline}\label{MTint2}
I[a,b]=\int\limits_{\R^{3}\times\CP^1\times\CP^1} \d^{3}x\;\la\lambda_1 \d\lambda_{1}\ra\;\la\lambda_{2}\d\lambda_{2}\ra\;\la\lambda_{1}\lambda_{2}\ra^{2} \\
\times \tr\!\left(b(x,\lambda_1)\,U_{X}(\lambda_{1},\lambda_{2})\,b(x,\lambda_{2})\,U_{X}(\lambda_{2},\lambda_{1})\right)\,,
\end{multline}
with the holomorphic frames serving to transport the insertions of $b$ between the two different insertion points on $X\cong\CP^1$.

The full minitwistor action is
\be\label{MTact1}
	S[a,b]=S_{m}[a,b]+\frac{g^2}{2}\,I[a,b]\,.
\ee
Note that the anti--monopole interaction term $I[a,b]$ is independent of the components $a_\perp$ and $b_\perp$ pointing along the fibres of $\PS\to\MT$, since these components wedge to zero in the perturbative expansion of \eqref{MTint2}. So on--shell, it remains true that $\bar{a}$ and $\bar{b}$ are independent of the fibre directions, and they can be considered to be pulled back from forms on $\MT$. However, the equations of motion no longer imply that the bundle over $\MT$ is holomorphic. In fact, one finds
\be\label{MTeom2}
	\cF^{0,2} = g^2\,\d^2 x_{(0,2)}\,\int_{X}\la\lambda'\,\d\lambda'\ra\,\la\lambda\,\lambda'\ra^2\, U_{X}(\lambda,\lambda')\,\bar{b}(x,\lambda')\,U_{X}(\lambda',\lambda)\,,
\ee
and 
\begin{multline}\label{MTeom3}
\bar{D} b= g^2\,\d^2 x^{\alpha\beta}_{(0,2)}\,\frac{\hat{\lambda}_{\alpha}\hat{\lambda}_{\beta}}{\la\lambda\,\hat{\lambda}\ra^4} \,\int\limits_{\CP^1\times\CP^1} \la\lambda'\d\lambda'\ra\,\la\lambda''\d\lambda''\ra\,\la\lambda'\hat{\lambda}\ra\,\la\lambda''\hat{\lambda}\ra\,\la\lambda'\lambda''\ra \\
\times\left[U_{X}(\lambda,\lambda')\,\bar{b}(x,\lambda')\,U_{X}(\lambda',\lambda),\,U_{X}(\lambda,\lambda'')\,\bar{b}(x,\lambda'')\,U_{X}(\lambda'',\lambda)\right]\,,
\end{multline}
where the 2-forms
\begin{equation*}
 \d x^{\alpha\beta}_{(0,2)}:=\d x^{\delta\gamma}\wedge\d x^{\sigma}_{\gamma}\,\frac{\lambda^{\alpha}\lambda^{\beta}\,\hat{\lambda}_{\delta}\hat{\lambda}_{\sigma}}{\la\lambda\,\hat{\lambda}\ra^2}\,, \qquad \d^{2}x_{(0,2)}:=\frac{\hat{\lambda}_{\alpha}\hat{\lambda}_{\beta}}{\la\lambda\,\hat{\lambda}\ra^2}\,\d^{2}x^{\alpha\beta}_{(0,2)}\,,
\end{equation*}
are projected to point in the anti-holomorphic directions of $\MT$. The two equations \eqref{MTeom2} and \eqref{MTeom3} are in fact equivalent to the field equations of YMH theory, written in the form \eqref{CSeom}; this follows from an argument similar to the one given for the four-dimensional Yang-Mills equations in~\cite{Mason:2005zm}. We will not show this in detail here, because in the next section we obtain the stronger result that the actions~\eqref{MTact1} and~\eqref{stact1} are in fact equivalent {\it off--shell}.

\medskip

Generalising the minitwistor action to $\cN=8$ SYMH$_3$ theory is straightforward. Encoding the field content into a single multiplet $\cA$ (defined initially on $\PS$), the resulting action takes the form:
\be\label{sMTact1}
	S[\cA]
	=\int_{\PS_s} \Omega_s \wedge\tr\left(\cA\wedge\d\cA+\frac{2}{3}\cA\wedge\cA\wedge\cA\right) + \frac{g^2}{2}\int_{\R^{3|8}}\d^{3|8}x\,\log\det\left(\dbar+\cA\right)\!|_{X}\,.
\ee
The first term here is a `partially holomorphic' Chern-Simons theory on $\PS$, which was shown to describe the supersymmetric Bogomolny sector in~\cite{Popov:2005uv}. The $\log\det(\dbar+\cA)|_{X}$ term can be understood perturbatively, via the expansion
\be\label{logdet}
\log\det\left(\dbar+\cA\right)|_{X}=\tr\!\left(\log\dbar|_{X}\right) + \sum_{n=2}^{\infty}\frac{1}{n}\int_{(\CP^1)^{\times n}} \tr\!\left(\dbar^{-1}|_{X}\cA(\lambda_1) \cdots\dbar^{-1}|_{X}\cA(\lambda_n)\right)\,,
\ee
where $\dbar|_{X}$ is the $\dbar$-operator along a minitwistor line. We can also write
\begin{multline}
 	\int\limits_{(\CP^1)^{\times n}} \hspace{-0.3cm}\tr\!\left(\dbar^{-1}|_{X}\cA(\lambda_1) \cdots\dbar^{-1}|_{X}\cA(\lambda_n)\right) \\
	=\int\limits_{(\CP^1)^{\times n}} \hspace{-0.3cm}\frac{\la\lambda_1\d\lambda_1\ra\,\cdots\,\la\lambda_n\d\lambda_n\ra}{\la\lambda_1\lambda_2\ra\la\lambda_2\lambda_3\ra\cdots\la\lambda_n\lambda_1\ra}\,\tr \left(\cA(\lambda_1)\,\cA(\lambda_2)\cdots\cA(\lambda_n)\right)\,, 
\end{multline}
using the Cauchy kernel on $\CP^1$. 

The action~\eqref{MTact1} is obviously very closely related to the twistor action
\be\label{N=4twistoract}
	S[\cA_4] = \int\limits_{\PT_s}\D^{3|4}Z\wedge\tr\left(\cA_4\wedge\dbar\cA_4+\frac{2}{3}\cA_4\wedge\cA_4\wedge\cA_4\right) + \frac{g^2_4}{2}\int\limits_{\R^{4|8}}\d^{4|8}x\,\log\det\left(\dbar+\cA_4\right)\!|_{X}
\ee
obtained in~\cite{Boels:2006ir} that describes $\cN=4$ SYM in four dimensions. The first term in~\eqref{N=4twistoract} is an integral over the $\cN=4$ twistor space $\PT_s\cong \cO(1)\otimes\C^{2|4}\to\CP^1$, where $\cA_4\in\Omega^{0,1}(\PT_s,{\rm End}(E))$, while the second term is an integral over the space of real Euclidean twistor lines $X\subset\PT_s$, corresponding to chiral $\cN=4$ superspace in four dimensions. Indeed, if $\cL_T(\cA_4)=0$ so that $\cA_4$ is invariant along the flow of the vector field $T= T^{\alpha\dot\alpha}\lambda_\alpha\del/\del\mu^{\dot\alpha}$ that reduces $\PT_s$ to $\MT_s$, then identifying
\be\label{sTred}
	\Omega_s = T\lrcorner\, \D^{3|4}Z\qquad\text{and}\qquad \d^{3|8}x = T\lrcorner\, \d^{4|8}x\,,
\ee
we obtain the minitwistor action \eqref{sMTact1} upon taking the symmetry reduction from $\PT_s$ along the integrable curves of $T$.


\subsection{Equivalence to space-time action}

We now establish the off--shell equivalence of the action~\eqref{sMTact1} with the $\cN=8$ SYMH action on $\R^3$. This of course also follows from its status as the symmetry reduction of the twistor action~\eqref{N=4twistoract} for $\cN=4$ SYM, but we prefer to show it here from scratch. We first choose a basis of forms on $\PS$ that is adapted to the description $\PS\cong \R^3\times\CP^1$. 
Define the forms
\be\label{formbasis}
	\bar{e}^0 \equiv \frac{\la\hat\lambda\,\d\hat\lambda\ra}{\la\lambda\,\hat\lambda\ra^2}
	\qquad\text{and}\qquad
	\bar{e}^\alpha \equiv \frac{ \d x^{\alpha\beta}\,\hat{\lambda}_{\beta}\,}{\la\lambda\,\hat{\lambda}\ra}
\ee
which are dual to the vector fields
\be
	\dbar_{0}\equiv \la\lambda\,\hat{\lambda}\ra\,\lambda_{\alpha}\frac{\partial}{\partial\hat{\lambda}_{\alpha}}
	\qquad\text{and}\qquad
	\dbar_\alpha \equiv \lambda^\beta\frac{\del}{\del x^{\alpha\beta}}\,,
\ee
respectively. We have chosen to normalize these forms and vector fields by powers of $\la\lambda\,\hat\lambda\ra$ such that they have no antiholomorphic weight.  Note also that
\be
	\Omega\wedge\bar{e}^0\wedge\bar{e}^\alpha\wedge\bar{e}_\alpha
	= \frac{\la\lambda\,\d\lambda\ra\wedge\la\hat{\lambda}\,\d\hat{\lambda}\ra}{\la\lambda\,\hat{\lambda}\ra^2}\wedge\d^3x\,,
\ee
so~\eqref{formbasis} span the directions of $\PS$ not involved in the holomorphic form $\Omega$.

In terms of the basis~\eqref{formbasis} we can expand the field $\cA$ as
\be\label{sMTfield}
	\cA(x,\lambda,\hat{\lambda},\theta\lambda)=\cA_{0}(x,\lambda,\hat{\lambda},\theta\lambda)\,\bar{e}^{0} + \cA_{\alpha}(x,\lambda,\hat{\lambda},\theta\lambda)\,\bar{e}^{\alpha}\,.
\ee
We now exploit the gauge redundancy of the action~\eqref{sMTact1} to impose the gauge
\be\label{hgauge}
	\dbar^*|_{X} \cA|_{X}=0
\ee
on every minitwistor line $X$, where $\dbar^*|_{X}$ is the adjoint of $\dbar|_X$ with respect to the standard Fubini--Study metric on $X\cong\CP^1$. (The action does not require any choice of metric except through this gauge--fixing term.) We also have $(\dbar\cA)|_X=0$ for trivial dimensional reasons, so in this gauge $\cA|_X$ is fixed to be a harmonic representative of the cohomology group 
\be
	H^{0,1}(X, {\rm End}(E)\otimes\cO_s)\cong \bigoplus_{n=0}^{4} H^{0,1}(X,{\rm End}(E)\otimes\cO_X(-n))\otimes \C^{\frac{4!}{n!(4-n)!}}\,,
\ee
where the right hand side gives the cohomology groups describing the component fields in the expansion of the supermultiplet $\cA = a + \cdots+\chi^4 b$. These cohomology groups vanish if $n<2$ so the component fields $a|_{X}$ and $\psi_a|_X$ vanish. Harmonic representatives for the remaining fields are~\cite{Woodhouse:1985id}
\be
	\varphi_{ab}|_X =  \phi_{ab}(x)\,\bar{e}^{0}\,,\qquad \tilde{\psi}^{a}|_X=2\,\frac{\widetilde{\Psi}^{a}_{\alpha}(x)\,\hat{\lambda}^{\alpha}}{\la\lambda\,\hat{\lambda}\ra}\,\bar{e}^{0}
	\qquad\text{and}
	\qquad
	b|_X = 3\,\frac{B_{\alpha\beta}(x)\,\hat{\lambda}^{\alpha}\hat{\lambda}^{\beta}}{\la\lambda\,\hat{\lambda}\ra^2}\,\bar{e}^0
\ee
where the fields $\{\phi_{ab},\tilde\Psi^a_{\alpha}, B_{\alpha\beta}\}$ can depend only on $x\in\R^3$. While the components of $\cA$ restricted to (real Euclidean) minitwistor lines are fixed, our gauge condition does not constrain the remaining components of $\cA$. Thus, in this gauge we have 
 \be\label{hgaugefields}
 \begin{aligned}
 	a&=a_{\alpha}(x,\lambda,\hat{\lambda})\,e^{\alpha}\,, \qquad\qquad &\psi_{a}&=\psi_{a\,\alpha}(x,\lambda,\hat{\lambda})\,e^{\alpha}\,,\\
	\varphi_{ab}&=\phi_{ab}(x)\,\bar{e}^{0}+\varphi_{ab\,\alpha}(x,\lambda,\hat{\lambda})\,e^{\alpha}\,, \qquad\qquad
	&\tilde{\psi}^{a}&=2\,\frac{\widetilde{\Psi}^{a}_{\alpha}(x)\,\hat{\lambda}^{\alpha}}{\la\lambda\,\hat{\lambda}\ra}\,\bar{e}^{0} +\tilde{\psi}^{a}_{\alpha}(x,\lambda,\hat{\lambda})\,e^{\alpha}\,,\\
	b&=3\,\frac{B_{\alpha\beta}(x)\,\hat{\lambda}^{\alpha}\hat{\lambda}^{\beta}}{\la\lambda\,\hat{\lambda}\ra^2}\,\bar{e}^0 + b_{\alpha}(x,\lambda,\hat{\lambda})\,e^{\alpha}\,.&&
\end{aligned}
 \ee
This gauge is not a complete gauge fixing on $\PS$, with the residual gauge freedom being smooth gauge transformations $\gamma$ which obey
\be
 	\dbar^*|_X \dbar_X\gamma(x,\lambda,\hat{\lambda})=0
\ee
and so are themselves harmonic on minitwistor lines. Since $\gamma$ is homogeneous of weight zero on $\CP^1$, by the maximum modulus principle it follows that such $\gamma(x,\lambda,\hat{\lambda})=\gamma(x)$. Thus, in harmonic gauge on $\PS$, the residual gauge freedom of the minitwistor action is just ordinary gauge transformations on $\R^3$. 

\medskip

We now evaluate the minitwistor action using the harmonic gauge fields~\eqref{hgaugefields}. Consider first the monopole contribution $S_{\mathrm{m}}[\cA]$. Performing the Grassmann integration over $\d^4\chi$ is straightforward, leaving
\begin{multline}\label{ste1}
 \int\limits_{\PS}\d^{3}x\wedge\omega\,\tr\left[ 3\,\frac{B_{\alpha\beta}(x)\,\hat{\lambda}^{\alpha}\hat{\lambda}^{\beta}}{\la\lambda\,\hat{\lambda}\ra^2}\left(\lambda^{\delta}\partial^{\gamma}_{\delta} a_{\gamma} +\frac{1}{2}[a^{\gamma},a_{\gamma}]\right)+\phi^{ab}\,\psi_{a\,\alpha}\,\psi_{b}^{\alpha} \right.\\
 +2\,\frac{\widetilde{\Psi}^{a}_{\alpha}(x)\,\hat{\lambda}^{\alpha}}{\la\lambda\,\hat{\lambda}\ra}\left(\lambda^{\beta}\partial^{\gamma}_{\beta}\psi_{a\,\gamma}+[a^{\gamma},\psi_{a\,\gamma}]\right) +\frac{\phi^{ab}}{2}\left(\lambda^{\alpha}\partial^{\beta}_{\alpha}\varphi_{ab\,\beta}+[a^{\beta},\varphi_{ab\,\beta}]\right) \\
 \left.-b^{\alpha}\dbar_{0}a_{\alpha}-\tilde{\psi}^{a\,\alpha}\dbar_{0}\psi_{a\,\alpha}+\frac{1}{2}\varphi^{ab}_{\alpha}\,\dbar_{0}\varphi_{ab}^{\alpha}\right]\,,
\end{multline}
where
\be\label{kahler}
	\omega\equiv\frac{\la\lambda\,\d\lambda\ra\wedge\la\hat{\lambda}\,\d\hat{\lambda}\ra}{\la\lambda\,\hat{\lambda}\ra^2}\,,
\ee
is the K{\"a}hler form on $\CP^1$. The field components $b_{\alpha}$ and $\tilde{\psi}^{a}_{\alpha}$ appear only in the third line of \eqref{ste1}. Integrating them out of the path integral  enforces the constraints
\be\label{ste2}
	\dbar_{0}a_{\alpha}=0\,, \qquad \dbar_{0}\psi_{a\,\alpha}=0\,,
\ee
so that $a_\alpha$ and $\psi_{a\,\alpha}$ must be globally holomorphic in $\lambda$. Accounting for the weight $-1$ of the basis form $\bar{e}^\alpha$, we see that $\psi_{a\,\alpha}$ is homogeneous of weight zero with respect to $\lambda$ (and $\hat{\lambda}$), whilst $a_\alpha$ is homogeneous of weight $+1$ in $\lambda$. Thus the second of these constraints implies that $\psi_{a\,\alpha}=\Psi_{a\,\alpha}(x)$, whilst the first constraint implies that $a_{\alpha}=\lambda^{\beta}\tilde{A}_{\alpha\beta}(x)$. Decomposing this $\tilde{A}_{\alpha\beta}$ into its symmetric and anti-symmetric parts, the result is
\be\label{ste3}
	a_{\alpha}(x,\lambda,\hat{\lambda})=\lambda^{\beta}\,A_{\alpha\beta}(x)-2\lambda_{\alpha}\,\Phi(x) \qquad\text{and}\qquad 
	\psi_{a\,\alpha}(x,\lambda,\hat{\lambda})=\Psi_{a\,\alpha}(x)\,,
\ee
where $A_{\alpha\beta}=A_{\beta\alpha}$ defines a gauge field on $\R^3$ and $\Phi(x)$ is the Higgs field.

Having solved these constraints, we can perform the path integral over the components $\varphi_{ab\,\alpha}$, which is a Gaussian with quadratic operator $\dbar_0$. The result of this path integration leaves an action
\begin{multline}\label{ste6}
 \int\limits_{\R^3\times\CP^1} \d^{3}x\wedge\omega\, \tr\!\left[ 3\,\frac{B_{\alpha\beta}\,\hat{\lambda}^{\alpha}\hat{\lambda}^{\beta}}{\la\lambda\,\hat{\lambda}\ra^2}\lambda^{\gamma}\lambda^{\delta} \left(\partial^{\kappa}_{\gamma} A_{\kappa\delta}+[A_{\gamma}^{\kappa},\,A_{\kappa\delta}] -2\D_{\gamma\delta}\Phi\right)+\phi^{ab}\,\Psi_{a\,\alpha}\,\Psi_{b}^{\alpha} \right. \\
 +2\,\frac{\widetilde{\Psi}^{a}_{\alpha}\,\hat{\lambda}^{\alpha}\lambda^{\beta}}{\la\lambda\,\hat{\lambda}\ra}\left(\D^{\gamma}_{\beta}\Psi_{a\,\gamma}+[\Phi,\,\Psi_{a\,\beta}]\right) +\frac{1}{4}\,[\phi^{ab},\,\Phi]\,[\Phi,\,\phi_{ab}]\\
 \left. +\frac{\phi^{ab}\,\lambda^{\alpha}\hat{\lambda}^{\beta}}{2\,\la\lambda\,\hat{\lambda}\ra}\left(\D^{\gamma}_{\alpha}\D_{\beta\gamma}\phi_{ab}+\partial_{\alpha\beta}[\Phi,\,\phi_{ab}]\right)\right]\,.
\end{multline}
At this point, the integral over the $\CP^1$ factor of $\PS\cong\CP^1\times\R^3$ can be performed using the rule~\cite{Boels:2006ir}
\be\label{Sdual}
\int_{\CP^1}\frac{\omega}{\la\lambda\,\hat{\lambda}\ra^{n}} \lambda^{\alpha_1}\cdots\lambda^{\alpha_n} S_{\alpha_1 \cdots\alpha_n}\,\hat{\lambda}_{\beta_1}\cdots\hat{\lambda}_{\beta_n} T_{\beta_1\cdots\beta_n} = \frac{1}{n+1}S_{\alpha_1 \cdots\alpha_n}\,T^{\alpha_1\cdots\alpha_n}\,,
\ee
which is a consequence of Serre duality on the Riemann sphere. The result is
\begin{multline}\label{ste7}
	 \int_{\R^3} \d^{3}x\;\tr\!
	 \left\{B_{\alpha\beta}\left(\partial^{\alpha\gamma} A^{\beta}_{\gamma} + [A^{\alpha\gamma},\,A^{\beta}_{\gamma}] -2\D^{\alpha\beta}\Phi\right) 
	 -\widetilde{\Psi}^{a}_{\alpha}\,\D^{\alpha\beta}\Psi_{a\,\beta} + \Psi_{a\,\alpha}\,\Phi\,\widetilde{\Psi}^{a\,\alpha} \phantom{\frac{1}{4}}\right.\\
	\left.+\Psi_{a\,\alpha}\Psi^{\alpha}_{b}\,\phi^{ab}+\frac{1}{4}\phi_{ab}\,\D_{\alpha\beta}\D^{\alpha\beta}\phi^{ab}+\frac{1}{4}[\phi_{ab},\Phi]\,[\Phi,\phi^{ab}]	\right\}\,.
\end{multline}
This is equal to the monopole action for $\cN=8$ SYMH theory on $\R^3$, up to a total derivatives which can be discarded.

\medskip

The calculation for the interaction term $I[\cA]$ follows similar lines. In harmonic gauge, the perturbative expansion of $\log\det(\dbar+\cA)|_{X}$ terminates at fourth-order (because $\cA_{0}$ goes like $\chi^2$ at leading order in this gauge), so there are relatively few terms to consider. These are further reduced by the requirement that the fermionic integral over $\d^{8}\theta$ must be saturated. As it turns out, only three such terms are present in the perturbative expansion: one at second order ($\sim b^2$), one at third order ($\sim\phi\widetilde{\psi}\widetilde{\psi}$), and one at fourth order ($\sim\phi^4$). We will only review the calculation for the second order term; the others follow a similar path. 

The relevant second-order contribution is
\begin{multline}\label{ste1*}
\int \d^{3|8}x \int\limits_{(\CP^1)^{\times2}}\frac{\la\lambda_1\,\d\lambda_1\ra}{\la\lambda_2\,\lambda_1\ra}b_{0}(x,\lambda_1,\hat{\lambda}_{1})\,\bar{e}^{0}_{1}\,\chi_{1}^{4}\,\frac{\la\lambda_2\,\d\lambda_2\ra}{\la\lambda_1\,\lambda_2\ra}b_{0}(x,\lambda_2,\hat{\lambda}_{2})\,\bar{e}^{0}_{2}\,\chi_{2}^{4}\\
=-9\int\d^{3|8}x \int\limits_{(\CP^1)^{\times2}}\frac{\omega_1\,\omega_2}{\la\lambda_1\,\lambda_2\ra^2}\frac{B_{\alpha\beta}\,\hat{\lambda}^{\alpha}_{1}\hat{\lambda}^{\beta}_{1}}{\la\lambda_{1}\,\hat{\lambda}_{1}\ra^2} (\theta^{a\,\gamma}\lambda_{1\,\gamma})^4\,\frac{B_{\delta\kappa}\,\hat{\lambda}^{\delta}_{2}\hat{\lambda}^{\kappa}_{2}}{\la\lambda_{2}\,\hat{\lambda}_{2}\ra^2} (\theta^{b\,\sigma}\lambda_{2\,\sigma})^4\,,
\end{multline}
where the incidence relations $\chi^{a}=\theta^{a\alpha}\lambda_{\alpha}$ have been used. It can be shown that
\be\label{fermint}
	\int \d^{8}\theta\,(\theta^{a\gamma}\lambda_{1\gamma})^4\,(\theta^{b\beta}\lambda_{2\beta})^4=\la\lambda_{1}\lambda_{2}\ra^{4}\,,
\ee
which enables the further reduction of the second-order contribution to:
\begin{multline}\label{ste2*}
-9\int\limits_{\R^3\times\CP^1\times\CP^1}\d^{3}x\,\frac{\omega_{1}\,\omega_{2}}{\la\lambda_{1}\,\hat{\lambda}_{1}\ra^2\,\la\lambda_{2}\,\hat{\lambda}_{2}\ra^2}\,\la\lambda_1\,\lambda_2\ra\,B_{\alpha\beta}\,\hat{\lambda}^{\alpha}_{1}\hat{\lambda}^{\beta}_{1}\,B_{\gamma\delta}\hat{\lambda}^{\gamma}_{2}\hat{\lambda}^{\delta}_{2} \\
=-\int_{\R^3}\d^{3}x\,B_{\alpha\beta}\,B^{\alpha\beta}\,,
\end{multline}
after making use of \eqref{Sdual}. This is precisely the $B^2$ contribution to the anti-monopole interactions for $\cN=8$ SYMH$_3$. The other two terms in the space-time action are generated by the third and fourth order contributions from the perturbative expansion of $\log\det$ in a similar fashion.


\section{Tree Amplitudes in YMH$_3$ Theory}
\label{sec:Amps}

For $\cN=4$ super-Yang-Mills theory in four dimensions, the perturbative expansion around the self--dual sector has its ultimate expression in the RSVW formula~\cite{Witten:2003nn,Roiban:2004yf}
\be\label{RSVW}
	\sum_{d=0}^\infty g_{(4)}^{2d} \int \d\tilde{\mu}^{(d)}_{\tilde{C}} \,\log\det(\dbar + \cA_4)|_{\tilde{C}}
\ee
where $\tilde{C}$ is a the image of a degree $d$ holomorphic map $Z:\CP^1\to\CP^{3|4}$ from a rational curve to $\cN=4$ twistor space $\CP^{3|4}$, while 
\be
	\d\tilde\mu^{(d)}_{\tilde{C}} = \frac{\d^{4|4\times(d+1)}Z}{{\rm vol \,GL}(2;\C)}\,,
\ee
is a top holomorphic form on the moduli space of all such maps, described in terms of homogeneous coordinates on the target and considered up to automorphisms of the source curve\footnote{It is easily checked that, like $\CP^{3|4}$ itself, the moduli space is a Calabi--Yau supermanifold. See also~\cite{Witten:2003nn,Movshev:2006py,Adamo:2012cd}}, and $\dbar+ \cA_4$ is a (0,1)--connection on a complex holomorphic bundle $E\to\CP^{3|4}$. Expanding in powers of the on-shell background field $\cA_4$, this formula is a generating functional for all tree amplitudes in $\cN=4$ SYM$_4$. The degree of the map indicates the grading of the scattering amplitude by NMHV degree, with a degree $d$ map corresponding to a N$^{d-1}$MHV tree amplitude. 

Since Yang-Mills-Higgs theory inherits its perturbative expansion around solutions of the Bogomolny equations from the MHV expansion of Yang-Mills theory in four dimensions, it is natural to ask if a similar connected prescription exists for the tree-level S-matrix of YMH theory in three dimensions. Indeed, a formula along these lines was given in~\cite{Cachazo:2013iaa} as a literal restriction of the RSVW formula to three--dimensional kinematics. In this section we present a new formula that is adapted to the minitwistor geometry appropriate for the three--dimensional theory.


\subsection{A connected prescription generating functional}

In three dimensions an on-shell gluon has only one polarization state, so we cannot hope to have any analogue of an `MHV' expansion for pure Yang-Mills theory. However, in YMH$_3$ theory it is natural to grade $n$-particle perturbative amplitudes according to how many of the external particles depart from solutions of the (linearised) Bogomlony equations.

On $\MT_s$, the amplitudes of $\cN=8$ SYMH$_3$ theory can be viewed as functionals of the on-shell supermultiplet $\cA$, given by \eqref{sPT1}. As in four dimensions, the fermionic expansion of this supermultiplet automatically keeps track of this three--dimensional `MHV' expansion, as noted in~\cite{Chiou:2005jn,Agarwal:2011tz}. In particular, any tree-level amplitude $\cM^{(0)}_{n}$ can be expanded as a polynomial in the fermionic components of the on-shell supermomenta $\{\chi^{a}_{i}\}$, starting with a term of order $4$ and truncating at order $4(n-2)$. The $k^{\mathrm{th}}$ term in this expansion is of order $4(k+2)$ and is identified as the N$^k$MHV superamplitude; if we project out all but the top and bottom components of the supermultiplet, this N$^k$MHV amplitude would contain $k+2$ external states that obey the linearised `anti--Bogomolny' equations.

\medskip

With this understanding, our formula is remarkably similar to~\eqref{RSVW}: we find that all amplitudes in $\cN=8$ SYMH theory in three dimensions are given by the generating functional
\be\label{connectedlogdet}
	\sum_{d=0}^\infty g^{2d} \int \d\mu^{(d)}_C \,\frac{1}{R(\lambda)} \,\log \det(\dbar+\cA)|_C\,.
\ee
As in the RSVW formula, $\dbar+\cA$ is a (0,1)-connection on a background complex holomorphic bundle, here over $\MT_s$.  $C$ is the image of a degree $d$ map 
\be\label{map1}
\begin{aligned}
	Z&:\CP^1\rightarrow\ \MT_s\,,\\
		&:\ [\sigma^\mathbf{a}]\  \mapsto \left[u(\sigma),\,\lambda_{\alpha}(\sigma),\,\chi^{a}(\sigma)\right]
\end{aligned}
\ee
from a Riemann sphere, described by homogeneous coordinates $[\sigma^\mathbf{a}] =[\sigma^{\mathbf{0}},\sigma^{\mathbf{1}}]$, to maximally supersymmetric minitwistor space. Explicitly, we have
\be\label{map2}
\begin{aligned}
	&u(\sigma)=u_{\mathbf{a}_1 \cdots\mathbf{a}_{2d}}\,\sigma^{\mathbf{a}_1}\cdots \sigma^{\mathbf{a}_{2d}}\,, \qquad 
	\lambda_{\alpha}(\sigma)=\lambda_{\alpha}^{\mathbf{a}_1 \cdots\mathbf{a}_{d}}\,\sigma_{\mathbf{a}_1}\cdots \sigma_{\mathbf{a}_{d}}\,,\\
 &\hspace{2.5cm}\chi^{a}(\sigma)=\chi^{a}_{\mathbf{b}_1 \cdots\mathbf{b}_{d}}\,\sigma^{\mathbf{b}_1}\cdots \sigma^{\mathbf{b}_{d}}\,.
\end{aligned}
\ee
Note that the polynomial $u(\sigma)$ has degree $2d$, since it scales with twice the weight of $\lambda_{\alpha}$ and $\chi^{a}$ on minitwistor space. 

The measure
\be
	\d\mu^{(d)}_C = \frac{\d^{2d+1}u \,\d^{2|4\times(d+1)}\lambda}{{\rm vol\,GL}(2;\C)}
\ee
is a top holomorphic form on the space of such maps. Note that with this measure, the moduli space is not canonically Calabi-Yau: under a rescaling
\be\label{scalingRSVW}
	 (u,\lambda_{\alpha},\chi^{a})\rightarrow (r^{2}u, r\lambda_{\alpha}, r\chi^{a})
\ee
of the homogeneous coordinates on $\MT_s$, one finds
\be
	\d\mu^{(d)}_C \to r^{2(2d+1)} r^{(2-4)(d+1)} \,\d\mu^{(d)}_C = r^{2d} \,\d\mu^{(d)}_C
\ee
so that $\d\mu^{(d)}_C$ has non--trivial scaling weight. This weight is compensated by the new ingredient $R(\lambda)$ which is the \emph{resultant} of the $\lambda$ components of the map $\lambda:\CP^1\rightarrow\CP^1$ ({\it c.f.}, \cite{Cachazo:2013zc}). By definition, this resultant is a homogeneous polynomial in the coefficients $\lambda_\alpha^{{\bf a}_1\cdots{\bf a}_d}$ of degree $d$ that vanishes if and only if the two polynomials $\lambda_\alpha(\sigma)$ have a simultaneous root; in other words, if there is some point $[\sigma_*]\in\CP^1$ for which $\lambda_\alpha(\sigma_*)=0$. Since $\lambda$ describes a map to $\CP^1$, this never occurs, so $1/R(\lambda)$ is nowhere singular. Since it is homogeneous of degree $d$, the resultant scales under~\eqref{scalingRSVW} as
\be
 	R(r\lambda)\rightarrow r^{2d}\,R(\lambda)\,.
\ee
so that the measure $\d\mu^{(d)}_C / R(\lambda)$ is in fact scale invariant and holomorphic. 

Clearly, the formula \eqref{connectedlogdet} shares many features with its four-dimensional avatar, the RSVW formula. Indeed, as we will show below, it can be seen as a straightforward symmetry reduction of the RSVW formula.

\medskip

We now demonstrate that~\eqref{connectedlogdet} does indeed yield the correct tree amplitudes. Expanding in powers of the background field $\cA$,  the $n$-particle tree-level N$^{d-1}$MHV superamplitude is given by
\be\label{cf1}
	\cM^{(0)}_{n,d}
	=\int \frac{\d\mu^{(d)}_C}{R(\lambda)}\,
	\prod_{i=1}^{n}\frac{(\sigma_i\d\sigma_i)}{(i\,i\!+\!1)}\ 
	\tr\left(\cA_1\,\cA_2\,\cdots\,\cA_n\right)\, ,
\ee
where $\cA_i = \cA(u(\sigma_i),\lambda(\sigma_i),\chi(\sigma_i))$, $(\sigma_i\d\sigma_i) = \epsilon_{\bf ab}\, \sigma^{\bf a}_i\d\sigma_i^{\bf b}$ and similarly $(i\,i\!+\!1) = \epsilon_{\bf ab}\, \sigma_i^{\bf a}\sigma_{i+1}^{\bf b}$, for $\epsilon_{\bf ab}$ the SL$(2;\C)$ invariant tensor on $\CP^1$. (The wavefunctions $\cA_i$ are weightless with respect to both the map components and the homogeneous coordinates $\sigma_i$ of each marked point on $\CP^1$, ensuring that the entire expression is well-defined projectively.) Choosing the $\cA_i$ to be minitwistor representatives of (super)-momentum eigenstates, we set
\be\label{pws}
\cA_{i}=\int \frac{\d t_i}{t_i}\,\bar{\delta}^{2}\!\left(\lambda_{i}-t_{i}\lambda(\sigma_i)\right)\,\exp\left[\im t_{i}^{2}\,u(\sigma_i)+\im t_{i}\eta_{i\,a}\,\chi^{a}(\sigma_i)\right]\,.
\ee
Via the Penrose transform, it is easy to see that such wavefunctions correspond to plane wave momentum eigenstate superfields
\be\label{pws2}
	\cA_{i}\rightarrow \exp\im\left(x^{\alpha\beta}\lambda_{i\,\alpha}\lambda_{i\,\beta} + \theta^{a\alpha}\eta_{i\,a}\lambda_{i\,\alpha}\right)\,,
\ee
on space-time, with on-shell supermomenta $\{\lambda_{i}\lambda_i,\,\lambda_{i}\eta_{i}\}$.

\medskip 

First, when $d=0$ the map $Z:\CP^1\to\MT_s$ is in fact constant, so $Z(\sigma)=Z$ and we take $R(\lambda)=1$ as standard. Thus, $\d\mu_C^{(0)} = \d u \,\d^2\lambda\,\d^4\chi / {\rm vol(GL(2;\C))}$, so using the ${\rm vol}\,SL(2;\C)$ factor to fix $\sigma_1,\sigma_2,\sigma_3$ to three arbitrary points, the three--point $d=0$ amplitude of~\eqref{cf1} becomes
\be\label{cf32}
\begin{aligned}
\cM^{(0)}_{3,0}
	&=\int \,\frac{\d u \,\d^2\lambda\,\d^4\chi}{ {\rm vol\,GL(2,\C)}}
	\prod_{i=1}^{3}\frac{(\sigma_i\d\sigma_i)}{(i\,i\!+\!1)}\ 
	\tr\left(\cA_1(\sigma_1)\,\cA_2(\sigma_2)\,\cA_3(\sigma_3)\right)\\
	&=\int \d u \,\la\lambda\,\d\lambda\ra\,\d^4\chi\,\tr\left(\cA_1\,\cA_2\,\cA_3\right)\,.
\end{aligned}
\ee
This is just the evaluation of the vertex of the action~\eqref{sMTact1} on three on-shell states. We have already demonstrated that this part of the action reduces to the first--order part of the space--time action corresponding to the (super-)Bogomolny equations, so it gives the same amplitudes. In fact, three particle momentum conservation $\lambda_1\lambda_1 + \lambda_2\lambda_2 +\lambda_3\lambda_3=0$ implies that $\la12\ra$, $\la23\ra$ and $\la31\ra$ all vanish, as also follows from the locality of the vertex in~\eqref{cf32}. ince there are no independent $\tilde\lambda_i$s in three dimensions, all three particle amplitudes must vanish even with complexified kinematics.

In fact, all three-point amplitudes vanish in three dimensions as a consequence of momentum conservation (even with complexified kinematics), so the 3-point $\overline{\mathrm{MHV}}$ is zero. However, one can still identify a meaningful $\overline{\mathrm{MHV}}$ pseudo-amplitude as the coefficient of the overall bosonic momentum conserving $\delta$-function. In~\cite{Lipstein:2012kd} this pseudo-amplitude was shown to take the form
\be\label{cfMHVbar}
	\cM^{(0)}_{3,0}=\delta^{3}\!\left(\sum_{i=1}^{3}\lambda_{i}\lambda_{i}\right) \times \frac{\delta^{0|4}(\eta_{1}\la 23\ra +\eta_{2}\la 31\ra + \eta_{3}\la 12\ra)}{\la12\ra\la23\ra\la31\ra}=0\,,
\ee

\medskip

It is also straightforward to show that \eqref{cf1} reproduces all of the MHV tree amplitudes of $\cN=8$ SYMH$_3$. Such amplitudes correspond to degree $d=1$ for the map to $\MT_s$, and the moduli integrations can be performed explicitly against the wavefunctions \eqref{pws}, leading to:
\be\label{cfMHV}
\cM^{(0)}_{n,1}=\delta^{3}\!\left(\sum_{i=1}^{n}\lambda_{i}\lambda_{i}\right)\,\delta^{8}\!\left(\sum_{i=1}^{n}\lambda_{i}\eta_{i}\right)\frac{1}{\la12\ra\,\la23\ra\cdots\la n-1\,n\ra\,\la n1\ra}\,,
\ee
which is the $n$-point MHV tree amplitude for $\cN=8$ SYMH$_3$ theory in the form obtained in~\cite{Lipstein:2012kd}.

\medskip

The strongest test of the formula's validity is factorization: locality and unitarity of $\cN=8$ SYMH$_3$ dictate that its tree amplitudes should have simple poles on multiparticle factorization channels, with no other singularities. Since \eqref{cf1} produces the correct MHV and $\overline{\mathrm{MHV}}$ seed amplitudes, BCFW recursion~\cite{Gang:2010gy} ensures that it is correct if it factorizes appropriately. 

Following studies of factorization for connected formulae of tree-level $\cN=4$ SYM and $\cN=8$ SUGRA in $d=4$~\cite{Skinner:2010cz,Cachazo:2012pz}, we can probe the factorization behaviour of \eqref{cf1} by looking at the limit where the underlying Riemann sphere degenerates. In a standard parametrization, $\Sigma_s\cong\CP^1$ degenerates in the $s\rightarrow 0$ limit to two Riemann spheres, $\Sigma_L$ and $\Sigma_R$ joined together at a node:
\begin{equation*}
 \lim_{s\rightarrow0}\Sigma_s = \Sigma_{L}\cup\Sigma_{R}\,.
\end{equation*}
If $\sigma^{\mathbf{a}}$ are the homogeneous coordinates on $\Sigma_{s}$, these are related to the natural coordinates on $\Sigma_L$ and $\Sigma_R$ by
\be\label{degcoord1}
\sigma_{L}^{\mathbf{a}}=\sigma^{\mathbf{0}}\left(\frac{\sigma^{\mathbf{1}}}{s},\,\sigma^{\mathbf{0}}\right)\,, \qquad \sigma_{R}^{\mathbf{a}}=\sigma^{\mathbf{1}}\left(\frac{\sigma^{\mathbf{0}}}{s},\,\sigma^{\mathbf{1}}\right)\,.
\ee
For simplicity, we can work with an affine coordinate $z$ on $\Sigma_s$ in the coordinate patch where $\sigma^{\mathbf{0}}\neq 0$, leading to
\be\label{degcoord2}
z_{L}=\frac{s}{z}\,, \qquad z_{R}=s\,z\,.
\ee
The origin of the two affine coordinates $z_L$, $z_R$ is the node $z_{\bullet}\in\Sigma_{L}\cap\Sigma_{R}$ where the two spheres are joined in the $s\rightarrow 0$ limit. 

We want to determine the behaviour of \eqref{cf1} in the $s\rightarrow 0$ limit. Without loss of generality, we assume that the $n$ original marked points are distributed into sets $L$, $R$ of size $n_{L}$ and $n_{R}$ on $\Sigma_L$ and $\Sigma_R$, respectively, in this limit. Likewise, the map into minitwistor space will degenerate into maps from $\Sigma_L$ and $\Sigma_R$ of degrees $d_L$ and $d_R$, respectively. These obey
\be\label{deg1}
n_{L}+n_{R}=n\,, \qquad d_{L}+d_{R}=d\,.
\ee
Straightforward calculations demonstrate the following small $s$ behaviour for the factors in the amplitude which depend only on the coordinates of $\Sigma_s$:
\be\label{degmeas1}
\frac{\prod_{i=1}^{n} \d z_{i}}{\mathrm{vol}\,\SL(2,\C)} = s^{n_{L}-n_{R}-4}\,\frac{\d s^{2}}{\prod_{i\in L}z_{L\,i}^2} \left(\frac{\prod_{j\in L\cup\{\bullet\}} \d z_{L\,j}}{\mathrm{vol}\,\SL(2,\C)}\right)  \left(\frac{\prod_{k\in R\cup\{\bullet\}} \d z_{R\,k}}{\mathrm{vol}\,\SL(2,\C)}\right)\,,
\ee
\be\label{degPT}
\prod_{i=1}^{n}\frac{1}{z_{i}-z_{i+1}} = s^{n_{R}-n_{L}+2}\,\prod_{i\in L}z_{L\,i}^{2}\,\prod_{j\in L\cup\{\bullet\}}\frac{1}{z_{L\,i}-z_{L\,i+1}}\,\prod_{k\in R\cup\{\bullet\}}\frac{1}{z_{R\,i}-z_{R\,i+1}}\,.
\ee
In particular, the measure and Parke-Taylor factors split into the relevant measures and Parke-Taylor factors on $\Sigma_L$ and $\Sigma_R$, up to overall factors of the parameter $s$. 

Now one must consider the portions of \eqref{cf1} which depend on the map to minitwistor space itself. For instance, the $\lambda_{\alpha}$ components of the map are written in the affine coordinate on $\Sigma_s$ as
\be\label{degmap1}
\lambda_{\alpha}(z)=\sum_{r=0}^{d}\lambda_{\alpha\,r}\,z^{r}\,.
\ee
Adapting this to the affine coordinate on each branch as $s\rightarrow 0$ allows $\lambda_{\alpha}(z)$ to be rewritten as
\be\label{degmap2}
 \lambda_{\alpha}(z)= z^{d_L}\left(\sum_{r=1}^{d_L}\lambda_{\alpha\,d_{L}-r}\,\frac{z_{L}^{r}}{s^r} + \lambda_{\alpha\,\bullet} + \sum_{t=1}^{d_R}\lambda_{\alpha\,d_L+t}\,\frac{s^{t}}{z_{L}^{t}}\right) 
\ee
\begin{equation*}
 = z^{d_L}\left(\sum_{r=1}^{d_L}\lambda_{\alpha\,d_{L}-r}\,\frac{s^{r}}{z_{R}^r} + \lambda_{\alpha\,\bullet} + \sum_{t=1}^{d_R}\lambda_{\alpha\,d_L+t}\,\frac{z_{R}^{t}}{s^{t}}\right)\,,
\end{equation*}
making the identification
\be\label{degmap3}
\lambda_{\alpha\,\bullet}:=\lambda_{\alpha\,d_L}\,.
\ee
Re-defining the map moduli according to
\be\label{degmap4}
\frac{\lambda_{\alpha\,d_L-r}}{s^r}\rightarrow \lambda_{\alpha\,r}\,, \qquad \frac{\lambda_{\alpha\,d_L+t}}{s^t}\rightarrow \lambda_{\alpha\,t}\,,
\ee
enables us to write the map in a way that is naturally adapted to the degeneration of $\Sigma_s$:
\be\label{degmap5}
\lambda_{\alpha}(z_L)=\lambda_{\alpha\,\bullet}+\sum_{r=1}^{d_L} \lambda_{\alpha\,r}\,z_{L}^{r} + \sum_{t=1}^{d_{R}} \lambda_{\alpha\,t}\,s^{2t}\,z_{L}^{-t} = \sum_{r=0}^{d_L}\lambda_{\alpha\,r}\,z_{L}^{r} + O(s^2)\,,
\ee
and
\be\label{degmap6}
 \lambda_{\alpha}(z_R)=\lambda_{\alpha\,\bullet}+\sum_{r=1}^{d_L} \lambda_{\alpha\,r}\,s^{2r}\,z_{R}^{-r} + \sum_{t=1}^{d_{R}} \lambda_{\alpha\,t}\,z_{R}^{t} = \sum_{t=0}^{d_R}\lambda_{\alpha\,t}\,z_{R}^{t} + O(s^2)\,.
\ee
A similar story holds for the $\chi^{a}(z)$ map components. The $u(z)$ portion of the map can also be written in a way that manifests the factorization:
\be\label{degmap7}
u(z_L)=u_{\bullet}+\sum_{r=1}^{2d_L}u_{r}\,z_{L}^{r} + O(s^2)= \sum_{r=0}^{2d_L}u_{r}\,z_{L}^{r}+O(s^2)\,,
\ee
\be\label{degmap8}
u(z_R)=u_{\bullet}+\sum_{t=1}^{2d_R}u_{t}\,z_{R}^{t} + O(s^2)= \sum_{t=0}^{2d_R}u_{t}\,z_{R}^{t}+O(s^2)\,,
\ee
although this requires a different re-scaling of the map moduli than \eqref{degmap4}:
\be\label{degmap9}
\frac{u_{2d_L-r}}{s^r}\rightarrow u_{r}\,, \qquad u_{2d_L}\rightarrow u_{\bullet}\,, \qquad \frac{u_{2d_L+t}}{s^t}\rightarrow u_{t}\,.
\ee
These choices present the map from $\Sigma_s$ to minitwistor space in a fashion that manifests factorization into degree $d_L$ and $d_R$ maps in the $s\rightarrow0$ limit.

These rescalings \eqref{degmap4} for the $\lambda_{\alpha}$ and $\chi^a$ moduli and \eqref{degmap9} for the $u$ moduli must be accounted for in the measure on these moduli. It is easy to see that the result is:
\be\label{degmeas2}
\d^{2d+1}u\,\d^{2(d+1)|4(d+1)}\lambda \rightarrow s^{d_{L}^2 + d_{R}^2}\,\d u_{\bullet}\,\d^{2|4}\lambda_{\bullet}\,\d^{2d_L}u_L\,\d^{2d_L|4d_L}\lambda_L\,\d^{2d_R}u_R\,\d^{2d_R|4d_R}\lambda_R\,.
\ee
The final place where $s$-scaling can appear in \eqref{cf1} is from the resultant $R(\lambda)$ in the denominator. To lowest order in $s$, one can show that~\cite{Cachazo:2012pz}
\be\label{degres}
R(\lambda)=s^{d_{L}^2 + d_{R}^2}\,R(\lambda_L)\,R(\lambda_R)\,,
\ee
where $R(\lambda_L)$, $R(\lambda_R)$ are the resultants of the degree $d_L$, $d_R$ maps $\lambda_{\alpha}(z_L)$ and $\lambda_{\alpha}(z_R)$ from $\Sigma_{L,R}$ to $\CP^1$ which emerge in the $s\rightarrow0$ limit.

Collecting all factors of $s$, one can now read off the behaviour of the formula as $s\rightarrow 0$:
\begin{multline}\label{degform}
\cM^{(0)}_{n,d}=\int \frac{\d s^2}{s^2}\,\frac{\d u_{\bullet}\,\d^{2|4}\lambda_{\bullet}}{\mathrm{vol}\,\C^*}\,\frac{\d^{2d_L}u_L\,\d^{2d_L|4d_L}\lambda_L}{\mathrm{vol}\,\SL(2,\C)}\,\frac{\d^{2d_R}u_R\,\d^{2d_R|4d_R}\lambda_R}{\mathrm{vol}\,\SL(2,\C)} \\
\times\prod_{i\in L\cup\{\bullet\}}\frac{\d z_{L\,i}}{z_{i}-z_{i+1}}\, \prod_{j\in R\cup\{\bullet\}} \frac{\d z_{R\,j}}{z_{j}-z_{j+1}}\, \prod_{k\in L}\cA_{k}\,\prod_{m\in R}\cA_{m}\,+ O(s^0)\,.
\end{multline}
In particular, the formula features a simple pole in $s^2$. To see that this corresponds to the simple pole in exchanged momentum, note that the total momentum inserted on $\Sigma_L$ in the $s\rightarrow 0$ limit is
\be\label{degmom1}
P_{L}^{\alpha\beta}=\sum_{i\in L}\lambda^{\alpha}_{i}\,\lambda^{\beta}_{i}\,.
\ee
Now, using the delta functions
\be\label{degmom2}
\prod_{i\in L}\bar{\delta}^{2}(\lambda_{i}-t_{i}\lambda(z_{L\,i}))\,,
\ee
which appear in \eqref{degform} through the wavefunction insertions on $\Sigma_L$, it follows that
\be\label{degmom3}
P_{L}^{\alpha\beta}=\sum_{i\in L}t_{i}^2\,\lambda^{\alpha}(z_{L\,i})\,\lambda^{\beta}(z_{L\,i}) = \sum_{i\in L}t_{i}^{2}\left(\sum_{r=0}^{d_L}\lambda^{\alpha}_{L\,r}\,z_{L}^{r}\right) \left(\sum_{t=0}^{d_L}\lambda^{\beta}_{L\,t}\,z_{L}^{t}\right)+O(s^2)\,.
\ee
Additionally, performing all of the intial $\d^{2d+1}u$ moduli integrals leads to a series of delta functions:
\be\label{degmom4}
\prod_{r=0}^{2d_{L}}\delta\!\left(\sum_{i\in L}t_{i}^{2}\frac{s^r}{z_{L\,i}^{r}}\right)\,.
\ee

On the support of these delta functions, the exchanged momentum obeys
\be\label{degmom5}
P_{L}^{\alpha\beta}=\lambda_{\bullet}^{\alpha}\,\lambda_{\bullet}^{\beta}\,\sum_{i\in L}t_{i}^2 + O(s^2)\,,
\ee
and therefore
\be\label{degmom6}
P_L^2 = O(s^2)\,.
\ee
Thus, the simple pole in $s^2$ which appears in \eqref{degform} can be identified with a simple pole of the form $P_{L}^{-2}$. So the degeneration limit $s\rightarrow 0$ corresponds precisely to the tree-level factorization channel we wanted to probe. Furthermore, the formula \eqref{cf1} has the desired simple pole in this channel. It is easy to see that these poles are the only such singularities in the formula, because the resultant $R(\lambda)$ is non-vanishing and all singularities of the Parke-Taylor factor correspond to factorization channels.

To complete the factorization argument, we must account for one set of additional moduli missing from the measure \eqref{degmeas2}. This can be done by inserting an auspicious factor of one into the formula:
\be\label{unity}
1=\int\limits_{\MT_s \times\MT_s}\frac{\d u_{*}\,\d^{2|4}\lambda_{*}}{\mathrm{vol}\,\C^*}\,\frac{\d u\,\d^{2|4}\lambda}{\mathrm{vol}\,\C^*}\,\frac{\d t}{t}\bar{\delta}(u-t^2 u_{\bullet})\,\bar{\delta}^{2|4}(\lambda-t\lambda_{\bullet})\,\frac{\d r}{r}\bar{\delta}(u-t^2 u_{*})\,\bar{\delta}^{2|4}(\lambda-t\lambda_{*})\,.
\ee
The measure over $\d u_{*}\,\d^{2|4}\lambda_{*}$ (along with its $\mathrm{vol}\C^*$ quotient) can now be incorporated into \eqref{degmeas2} to give the full factorized measure, while the new delta functions and scale integrals define state insertions at the node on either side of the factorization channel:
\be\label{intstate}
\cA_{L\,\bullet}=\int\frac{\d t}{t}\,\bar{\delta}(u-t^2 u_{\bullet})\,\bar{\delta}^{2|4}(\lambda-t\lambda_{\bullet})\,, \quad \cA_{R\,\bullet}=\int\frac{\d s}{s}\,\bar{\delta}(u-s^2 u_{*})\,\bar{\delta}^{2|4}(\lambda-s\lambda_{*})\,.
\ee
The residue of the formula on the simple pole in exchanged momentum is then
\be\label{factorized}
\int_{\MT_s} \frac{\d u\,\d^{2|4}\lambda}{\mathrm{vol}\,\C^*}\,\cM^{(0)}_{n_{L}+1,d_L}\!\left(\{\lambda_i,\eta_i\}_{i\in L};\, u,\lambda,\chi\right)\;\cM^{(0)}_{n_R+1,d_R}\!\left(u,\lambda,\chi;\,\{\lambda_{j},\eta_j\}_{j\in R}\right)\,.
\ee
The remaining integral over minitwistor space is simply the sum over on-shell states in $\cN=8$ SYMH$_3$ flowing through the cut.


\subsection{Relation to the RSVW formula}

We have already remarked on the close similarity between the RSVW formula for tree-level scattering in $\cN=4$ SYM$_4$ and the formula \eqref{cf1} for tree-level scattering in $\cN=8$ SYMH$_3$. This similarity is more than heuristic: the $\cN=8$ SYMH$_3$ formula can be viewed as a symmetry reduction of the RSVW formula itself. By expanding \eqref{RSVW}, the $n$-point N$^{d-1}$MHV tree amplitude of $\cN=4$ SYM$_4$ is given by
\be\label{RSVW1}
\widetilde{\cM}^{(0)}_{n,d}=\int \d\tilde{\mu}^{(d)}_{\tilde{C}}\,\,
	\prod_{i=1}^{n}\frac{(\sigma_i\d\sigma_i)}{(i\,i\!+\!1)}\ 
	\tr\left(\widetilde{\cA}_1\,\widetilde{\cA}_2\,\cdots\,\widetilde{\cA}_n\right)\,,
\ee
where $\widetilde{\cA}_i$ are (linearised) insertions of the $\cN=4$ SYM$_4$ on twistor space. Our claim is that $\widetilde{\cM}_{n,d}^{(0)}$ is reduced to $\cM^{(0)}_{n,d}$ upon replacing the twistor wavefunctions with minitwistor wave functions, $\widetilde{\cA}_{i}\rightarrow\cA_i$, and taking the symmetry reduction of the measure:
\be\label{RSVW2}
\mathcal{T}^{(d)}\lrcorner\,\d\tilde{\mu}^{(d)}_{\tilde{C}}=\frac{\d\mu^{(d)}_{C}}{R(\lambda)}\,.
\ee
This reduction is defined by taking the vector $\mathcal{T}^{(d)}$ on the moduli space of maps from $\CP^{1}\rightarrow\PT_s$ to be
\be\label{RSVW3}
 \mathcal{T}^{(d)}:=T^{\alpha\dot\alpha}\,\lambda_{\alpha}^{\mathbf{a}_1 \cdots\mathbf{a}_{d}}\,\frac{\partial}{\partial\mu^{\dot\alpha\,\mathbf{a}_1\cdots\mathbf{a}_d}}\,,
\ee
and the moduli of the $u$-component of the map $\CP^1\rightarrow\MT_s$ to be
\be\label{RSVW4}
u^{\mathbf{a}_1 \cdots\mathbf{a}_{2d}}=\lambda_{\alpha\phantom{\dot\alpha}}^{(\mathbf{a}_{1}\cdots\mathbf{a}_{d}}\,\mu^{\mathbf{a}_{d+1}\cdots\mathbf{a}_{2d})}_{\dot\alpha}\,T^{\alpha\dot\alpha}\,,
\ee
in terms of the moduli of the map to twistor space, so that $u(\sigma) = [\mu(\sigma)|T|\lambda(\sigma)\ra$.

The non-trivial part of the claim is the relationship between the measures on the map moduli \eqref{RSVW2}. Since both sides of \eqref{RSVW2} are weightless, the scaling weights (with respect to both map moduli and the coordinates on the Riemann sphere) match. Further, it is easy to see that the mass dimensions on both sides of the reduction match. Using the mass dimensions
\be\label{massd}
[\lambda]=\frac{1}{2}\,, \quad [\mu]=-\frac{1}{2}\,, \quad [u]=0\,, \quad [T]=1\,,
\ee
it is straightforward to see that $[\d\tilde{\mu}^{(d)}_{\tilde{C}}]=+1$, while $[\d\mu^{(d)}_{C}]=\frac{2d+2}{2}=d+1$. The resultant makes up the difference on the minitwistor side, since $[R(\lambda)]=d$.

\medskip

The relationship \eqref{RSVW2} can also be seen explicitly for low degree maps. When $d=0$, the relationship is precisely the reduction from $\PT_s$ to $\MT_s$ given by \eqref{sTred}. At $d=1$, we can compute directly:
\be\label{d1red1}
\begin{aligned}
\mathcal{T}^{(1)}\lrcorner\,\d\tilde{\mu}^{(1)}_{\tilde{C}}&=T^{\alpha\dot\alpha}\,\lambda^{\rm{\bf a}}_{\alpha}\,\frac{\partial}{\partial\mu^{\dot\alpha\,\mathbf{a}}}\,\lrcorner\,\frac{\d^{2}\mu^{{\bf0}}\wedge\d^{2}\mu^{{\bf1}}\wedge\d^{2|4}\lambda^{\bf 0}\wedge\d^{2|4}\lambda^{\bf 1}}{\mathrm{vol}\,\GL(2,\C)} \\
&= \frac{\d u^{\mathbf{00}}\wedge\d u^{\mathbf{01}}\wedge\d u^{\mathbf{11}}}{\la\lambda^{\mathbf{0}}\,\lambda^{\mathbf{1}}\ra}\,\frac{\d^{2|4}\lambda^{\bf 0}\wedge\d^{2|4}\lambda^{\bf 1}}{\mathrm{vol}\,\GL(2,\C)}\,,
\end{aligned}
\ee
using the identifications \eqref{RSVW3}, \eqref{RSVW4}. Sure enough, $\la\lambda^{\mathbf{0}}\,\lambda^{\mathbf{1}}\ra$ is precisely the resultant for the $\lambda$-components of the map when $d=1$. At $d=2$ we have $u^{\bf abcd} = [\mu^{({\bf ab}}|T|\lambda^{\bf cd)}\ra$ and
\be
	\mathcal{T}^{(2)} = 
	T^{\alpha\dot\alpha}\left(\lambda^{\bf00}_\alpha\frac{\del}{\del\mu^{\dot\alpha{\bf00}}} 
	+ \lambda^{\bf01}_\alpha\frac{\del}{\del\mu^{\dot\alpha{\bf01}}}
	+\lambda^{\bf11}_\alpha\frac{\del}{\del\mu^{\dot\alpha{\bf11}}}\right)\,.
\ee
Therefore we have
\be\label{d2red1}
\begin{aligned}
	\d^5u &= [\d\mu^{\bf 00}|T|\lambda^{\bf 00}\ra\wedge\left([\d\mu^{\bf00}|T|\lambda^{\bf01}\ra + [\d\mu^{\bf01}|T|\lambda^{\bf00}\ra\right)\\
	&\qquad \wedge\left([\d\mu^{\bf 00}|T|\lambda^{\bf 11}\ra + [\d\mu^{\bf 01}|T|\lambda^{\bf01}\ra + [\d\mu^{\bf11}|T|\lambda^{\bf00}\ra\right)\\
	&\qquad\qquad\wedge\left([\d\mu^{\bf01}|T|\lambda^{\bf11}\ra + [\d\mu^{\bf11}|T|\lambda^{\bf01}\ra\right)\wedge[\d\mu^{\bf11}|T|\lambda^{\bf11}\ra\quad &\text{mod $\d\lambda$}&\\
	&=\left(\la\lambda^{\bf 00}\lambda^{\bf 01}\ra\la\lambda^{\bf 01}\lambda^{\bf11}\ra - \la\lambda^{\bf00}\lambda^{\bf11}\ra^2\right)
	 \,\mathcal{T}^{(2)}\lrcorner \left(\d^2\mu^{\bf00}\wedge\d^2\mu^{\bf01}\wedge\d^2\mu^{\bf11}\right)\quad&\text{mod $\d\lambda$}&,
\end{aligned}
\ee
where in the first equality we neglect terms which give zero when wedged against $\d^{2}\lambda^{\bf00}\wedge\d^2\lambda^{\bf01}\wedge\d^2\lambda^{\bf11}$, and the second equality follows by repeated use of the Schouten identity and recalling $T^2=1$. The expression $\la\lambda^{\bf 00}\lambda^{\bf 01}\ra\la\lambda^{\bf 01}\lambda^{\bf11}\ra - \la\lambda^{\bf00}\lambda^{\bf11}\ra$ is exactly the resultant of the $d=2$ map, verifying~\eqref{RSVW2} in this degree 2 case. Higher degree cases follow similarly.


\section{Discussion}

In this paper, we have presented a new minitwistor action describing YMH theory in three dimensions, and also its maximally supersymmetric completion. We showed how this action reduces to the standard space--time action. The most obvious question is to understand how to perform perturbation theory using this action, obtaining a Feynman diagram expansion analogous to the MHV diagrams that follow from the twistor action for $\cN=4$ SYM in four dimensions~\cite{Boels:2006ir,Boels:2007qn,Adamo:2011cb}. It would also be interesting to construct an amplitude / super Wilson loop duality in three dimensions. While this might again be expected to mimicking the twistor approach of~\cite{Mason:2010yk,Bullimore:2011ni,Adamo:2011pv}, a significant difference would appear to be that in the three--dimensional case, even non--null separated space-time points correspond to intersecting minitwistor lines. Thus the minitwistor image of a piecewise null polygon appears to have many more `accidental' self--intersections, whose role in the Wilson loop would need to be understood. 

In 4d, this property was closely tied to dual conformal symmetry and the amplitudes of 3d $\cN=8$ SYM were shown to be dual conformal covariant in~\cite{Lipstein:2012kd}. Furthermore, dual conformal symmetry was demonstrated for the ABJM theory in~\cite{Huang:2010qy,Bargheer:2010hn}, and an amplitude/Wilson loop duality was found at 4-points in~\cite{Chen:2011vv}.

We have also presented a connected prescription formula for all tree amplitudes in (supersymmetric) YMH theory, demonstrating its correctness from $\overline{\rm MHV}$ and MHV examples, and from checking its properties under factorization. As with the RSVW formula in four dimensions, this expression cries out for an understanding in terms of a minitwistor string theory, perhaps along the lines of~\cite{Chiou:2005jn}. A worldsheet model that gauges the action of $T^{\dot\alpha\alpha}\lambda_\alpha\, \del/\del\mu^{\dot\alpha}$ would seem to be a good starting--point, though it also seems inevitable that any such model will in fact describe YMH theory coupled to some version of Einstein--Weyl gravity, this being the dimensional reduction of conformal gravity in four dimensions~\cite{Jones:1985pla}.

Perhaps most ambitiously, $3d$ YMH theory is also the arena for Polyakov's beautiful model of confinement through monopole condensation~\cite{Polyakov:1976fu}. Given the close relation between monopoles and the local part of the minitwistor action, it would be very interesting to understand how this occurs from the perspective presented here.

\acknowledgments

TA is supported by an Imperial College Junior Research Fellowship. DS is supported in part by a Marie Curie Career Integration Grant (FP/2007-2013/631289). JW is supported by an STFC studentship. This work has been partially supported by STFC consolidated grant ST/P000681/1.

\bibliography{YMH}
\bibliographystyle{JHEP}

\end{document}